\documentclass[final,3p,times]{elsarticle} % 12pt,

\usepackage{graphicx}
\usepackage{epsfig}
\usepackage{epstopdf}
\usepackage{amssymb}
\usepackage{amsthm}
\usepackage{bm}

\usepackage[colorlinks=true,citecolor=blue,linkcolor=black]{hyperref}
\usepackage{color}
\usepackage{amsmath}
\usepackage{multirow}
\usepackage[justification=centering]{caption}
\usepackage{subcaption}
\usepackage{color}
\newcommand{\norm}[1]{\left\lVert#1\right\rVert}
\usepackage[algoruled,boxed,lined]{algorithm2e}

\usepackage{booktabs} 
\journal{International Journal of Geomechanics}

\begin{document}

\begin{frontmatter}



\title{A stabilized total-Lagrangian SPH method for large deformation and failure in geomaterials}
\author[ess,vt]{Md Rushdie Ibne Islam}
\author[ess,boku]{Chong Peng\corref{bokuemail}}
\ead{pengchong@boku.ac.at}

\cortext[bokuemail]{Corresponding Author}
\address[ess]{ESS Engineering Software Steyr GmbH, Berggasse 35, 4400 Steyr, Austria}
\address[vt]{Department of Biomedical Engineering and Mechanics, Virginia Polytechnic Institute and State University, Blacksburg, VA 24061, USA}
\address[boku]{Institut f{\"{u}}r Geotechnik, Universit{\"{a}}t f{\"{u}}r Bodenkultur, Feistmantelstrasse 4, 1180 Vienna, Austria}

\begin{abstract}
Conventional smoothed particle hydrodynamics based on Eulerian kernels (CESPH) is widely-used in large deformation analysis in geomaterials. Despite being popular, it suffers from tensile instability and rank-deficiency; thus, it needs several numerical treatments to be stable. In this work, we present a stabilized total-Lagrangian SPH method (TLSPH), which is inherently free of tensile instability.  A stiffness-based hourglass control algorithm is employed to cure the hourglass mode caused by rank-deficiency. Periodic update of reference configuration is used in simulations to allow TLSPH to model large deformation and post-failure flow in geomaterials. Several numerical examples are presented to show the performance of the stabilized TLSPH method. The comparison between TLSPH and CESPH are discussed. The influences of hourglass control and configuration update are also discussed and shown in the numerical examples. It is found that the presented stabilized TLSPH is robust and can model large deformation and plastic flows in geomaterials. Particularly, the stabilized TLSPH delivers accurate and smooth stress results.
\end{abstract}


\end{frontmatter}


\section{INTRODUCTION}\label{sec-1}
In recent years, meshless Lagrangian particle-based methods, e.g. smoothed particle hydrodynamics (SPH) \cite{monaghan2012smoothed}, reproducing kernel particle method (RKPM) \cite{chen2017meshfree}, material point method (MPM) \cite{bardenhagen2004generalized}, become popular in the modeling of geotechnical problems involving large deformation and post-failure material flow. The reason is that these methods completely or partially avoid the use of mesh; thus, they can conveniently model problems with large deformation, strong material distortion, moving boundaries, and separation of material. Among them, SPH is a well-established truly meshless method having wide application in astrophysics, computational fluid dynamics, and solid mechanics \cite{monaghan2012smoothed}. It was initially used to simulate seepage failure and liquefaction by Maeda et al. \cite{maeda2006development} and Naili et al. \cite{naili20052d}. Then Bui and coworkers \cite{bui2008lagrangian,bui2011slope,bui2013improved} presented a comprehensive SPH framework for the modeling of large deformation and post-failure phenomena in geomaterials. Since then, SPH has been widely used in geotechnical applications such as large deformation \cite{bui2014novel}, granular flow \cite{chambon2011numerical,peng2016unified}, soil-structure interaction \cite{wang2014frictional,zhan2019three}, among others.

The previous research and applications of SPH in geotechnical engineering usually employ the conventional SPH, where the search of neighboring particles and computation of kernel functions are based on the deformed current configuration. This type of SPH kernel is named Eulerian kernel in the sense that particles can enter and exit its support domain. Consequently, the commonly-used SPH is termed as conventional Eulerian SPH (CESPH) hereafter. When applied to solid mechanics and geomechanics, the CESPH method has a number of deficiencies which reduce its accuracy and stability \cite{ganzenmuller2016hourglass}. Among them, the tensile instability \cite{swegle1995smoothed} and hourglass mode \cite{dyka1997stress} are the main issues restricting the application of CESPH in solid mechanics. Several numerical treatments were proposed to cure these instabilities. For tensile instability, Monaghan \cite{monaghan2000sph} and Gray et al. \cite{gray2001sph} developed the artificial pressure/stress method, where repulsive forces are introduced if there is tensile pressure/stress. The repulsive forces prevent particles from approaching too close to each other hence hinder the formation of particle clumps and artificial cracks. For hourglass mode, Dyka et al. \cite{dyka1997stress} identified that it is caused by rank-deficiency due to insufficient system integration. Accordingly, they proposed to use additional stress integration points to stabilize CESPH computations. Most geotechnical CESPH simulations employ artificial pressure/stress to alleviate the tensile instability \cite{bui2008lagrangian,bui2011slope,bui2014novel,peng2019loquat}, whereas the hourglass mode attracts less attention. 

Later, the root of tensile instability was identified by Belytschko et al. \cite{belytschko2000unified} to be the employment of Eulerian kernel function and the updated-Lagrangian formulation. Consequently, a total-Lagrangian SPH (TLSPH) was proposed and proved to be free of tensile instability \cite{bonet2002alternative,vignjevic2006sph}. As the Lagrangian kernel has nothing to do with the rank-deficiency, TLSPH still suffers from hourglass mode. Consequently, the path to a robust SPH method is to combine TLSPH with stress points or hourglass control algorithm. However, the stress point method drastically increases the computational cost and a clear rule on how to place and move the stress points is missing, which render it less attracting. Recently, an hourglass control algorithm was proposed for TLSPH based on the hourglass control mechanism used by finite elements with reduced integration \cite{ganzenmuller2016hourglass,ganzenmuller2015hourglass}. It is proved that with the hourglass control TLSPH has long term stability, free of tensile instability and hourglass mode, has quadratic convergence rate, and gives rise to accurate stress results\cite{ganzenmuller2015hourglass,zhan2019stabilized}. The last one is a desirable feature which is not yet achieved by other SPH variants in solid mechanics.

Although proved to be a reliable numerical method, TLSPH in its original form is unsuitable for large deformation and failure analysis in geomaterials. The total-Lagrangian formulation cannot handle large deformations and plastic flows, in which the physical connections of materials change under distortion, leading to unphysical deformation gradient and negative Jacobian in the total-Lagrangian formulation. In this work, we employ TLSPH to the modeling of large deformation and post-failure flow in geomaterials. To overcome the shortcomings of the original TLSPH, configuration update is employed with clearly defined criterion. The hourglass control algorithm from Ganzenmu\"uller \cite{ganzenmuller2015hourglass} is used to cure the hourglass mode. It is shown in the work that the stabilized TLSPH with configuration update can accurately model large deformation in geomaterials. Additionally, the comparison between the CESPH and TLSPH is carried out. The influences of hourglass control and configuration update are also discussed.

\section{GOVERNING EQUATIONS AND CONSTITUTIVE MODEL}
\subsection{Governing equations}
For the modeling of large deformation and failure in geomaterials, the governing equations consisting of mass and momentum conservations are written in the following Lagrangian forms in the current configuration

\begin{equation}\label{eq1}
    \dfrac{\mathrm{d}\rho}{\mathrm{d}t}=-\rho\nabla\cdot\bm v
\end{equation}

\begin{equation}\label{eq2}
\dfrac{\mathrm{d}\bm v}{\mathrm{d}t}=\dfrac{1}{\rho}\nabla \bm\sigma + \bm g
\end{equation}
where $\rho$, $\bm v$ and $\bm g$ are the density, velocity, and body force per unit mass. $\bm\sigma$ and $\nabla$ are the Cauchy stress tensor and divergence/gradient operator, respectively. $\mathrm{d}(\cdot)/\mathrm{d}t$ denotes the material derivative. Note that the above governing equations are written in respect with the current configuration $\bm x=\varphi(\bm X)$, where $\bm X$ is the coordinate in the initial configuration, and $\varphi$ is a mapping between the initial and current configurations. 

The Cauchy stress $\bm \sigma$ can be decomposed into two parts: deviatoric stress $\bm s$ and isotropic hydrostatic pressure $p$

\begin{equation}
    \bm \sigma = \bm s - p \bm I
\end{equation}
where $\bm I$ is a identity matrix. The pressure $p$ is defined as $p = -(\sigma_{xx} + \sigma_{yy} + \sigma_{zz} )/3$.

\subsection{Constitutive model}
The elasto-plastic Drucker-Prager (DP) model is employed to model the mechanical responses of geomaterials. The DP model has the following yield surface
\begin{equation}
f = \sqrt{J_2} - k_\phi p - k_c
\end{equation} where $J_2 = \bm s : \bm s / 2$ is the second invariant of the deviatoric stress $\bm s$, $k_\phi$ and $k_c$ are two constitutive parameters, which can be related to cohesion $c$ and frictional angle $\phi$ as

\begin{equation}
k_\phi = \frac{3\tan\phi}{\sqrt{9 + 12 \tan^2\phi}}
\end{equation}

\begin{equation}
k_c = \frac{3c}{\sqrt{9 + 12 \tan^2\phi}}
\end{equation}

In this work, the following plastic potential function with a non-associated plastic flow rule is employed
\begin{equation}
g^s = \sqrt{J_2} - k_{\varphi} p
\end{equation} where $k_\varphi$ is a model parameter related to the dilatancy angle $\varphi$
\begin{equation}
   k_\varphi = \frac{3 \tan\varphi}{\sqrt{9 + 12 \tan^2 \varphi}}
\end{equation}

In computations, a predicted elastic stress $\bm \sigma^*$ is first calculated assuming that there is no plastic effect

\begin{equation}
   \bm \sigma^* = \bm \sigma^t + \left(\bm \omega \cdot \bm \sigma^t - \bm \sigma^t  \cdot \bm \omega + 2 G \dot{\bm e} + K \epsilon_v \bm I\right) \Delta t
\end{equation}
where $\bm \sigma^t$ is the stress from the previous time step, $\bm \omega$ is the spin tensor, $\dot{\bm e} = \dot{\bm \varepsilon} - \dot{\varepsilon_v} \bm I$ is the deviatoric strain rate component, $\dot{\varepsilon_v} = (\dot{\varepsilon}_{xx} + \dot{\varepsilon}_{yy} + \dot{\varepsilon}_{xx})/3$. $G$, $K$ and $\Delta t$ are the elastic bulk modulus, the shear modulus, and the time step, respectively. Note that the Jaumann stress rate is used to ensure the objectivity in simulations with large deformation and rotation. The predicted stress $\bm \sigma^*$ is then subjected to plastic correction. If the predicted stress $\bm \sigma^*$ is inside the yield surface, it is accepted as the correct stress; otherwise, a plastic correction should be employed. In the plastic correction, the incremental plastic strain is obtained as

\begin{equation}
   \Delta \bm \epsilon^p = \Delta \lambda \frac{\partial g^s}{\partial \bm \sigma}
\end{equation} where the plastic multiplier increment $\Delta \lambda$ is calculated as

\begin{equation}
   \Delta \lambda = \frac{f \left(\bm \sigma^* \right)}{G + K k_\phi k_\varphi}
\end{equation} Consequently, the corrected stress at the current time step is evaluated as

\begin{equation}
   \bm s^{t + \Delta t} = \frac{\sqrt{J_2^*} - G \Delta \lambda}{\sqrt{J_2^*}} \bm s^*
\end{equation}

\begin{equation}
   p^{t + \Delta t} = p^* + K k_\varphi \Delta \lambda
\end{equation}

\begin{equation}
   \bm \sigma^{t + \Delta t} = \bm s^{t + \Delta t} - p^{t + \Delta t} \bm I
\end{equation}
where $\bm s^*$, $J_2^*$, and $p^*$ represents the predicted variables corresponding to the predicted stress $\bm \sigma^*$.

\section{CONVENTIONAL SPH FOR LARGE DEFORMATION ANALYSIS}\label{sec-sph}
\subsection{Eulerian kernel-based SPH approximation}
In SPH, the problem domain is represented by particles carrying field variables and moving with the material. The approximation of field functions and their derivatives are based on SPH kernel functions. In conventional SPH simulations, the kernel functions are defined in the current configuration with support domains of fixed shape. This conventional kernel function is termed as Eulerian kernel because particles can enter and exit a support domain as the material deforms. Note that although the approximation is based on Eulerian kernels, the conventional SPH is still a Lagrangian method because the particles move with the material. In this section, we briefly introduce the conventional Eulerian kernel-based SPH (CESPH) for large deformation analysis.

In CESPH, a field variable $f(\bm x)$ is approximated in the integral form

\begin{equation}
   <f(\bm x)> = \int_\Omega f(\bm x') W(\bm x - \bm x', h) \mathrm{d} \bm x'
\end{equation}
where $<\cdot>$ denotes the SPH integral approximation, $\Omega$ is the support domain of the kernel function $W(\bm x - \bm x',h)$, where $|| \bm x - \bm x'||$ is the distance in the current configuration and $h$ is the smoothing length. In this work, the Wendland C2 kernel function is used \cite{wendland1995piecewise}

\begin{equation}\label{kernel}
    W(q, h)=\alpha_d 
\begin{cases}
    (q + 0.5) (2 - q)^4, & \text{if } q\le 2\\
    0,                                & \text{otherwise}
\end{cases}
\end{equation}
where $\alpha_d=7/(32 \pi h^2)$ is a normalization factor, and $q=|| \bm x - \bm x'||/h$ is the normalized distance. 

Similarly, the derivative of the function $f(\bm x)$ can be approximated as

\begin{equation}
   <\nabla f(\bm x)> = -\int_\Omega f(\bm x') \nabla_{\bm x'} W(\bm x - \bm x', h) \mathrm{d} \bm x'
\end{equation}
where $\nabla_{\bm x'} W(\bm x - \bm x', h)$ denotes the derivative of the kernel function evaluated at $\bm x'$.

The discrete form of the SPH approximation equations can be written as the summation of contributions from all SPH particles in the support domain
\begin{equation}
   f(\bm x_i) = \sum_j f(\bm x_j) W(\bm x_i - \bm x_j, h) \frac{m_j}{\rho_j}
\end{equation}

\begin{equation}
   \nabla f(\bm x_i) = -\sum_j \left[f(\bm x_i) - f(\bm x_j) \right] \nabla_{\bm x_i} W(\bm x_i - \bm x_j, h) \frac{m_j}{\rho_j}
\end{equation} where $f(\bm x_i)$ and $f(\bm x_j)$ are the values of field variable at particles $i$ and $j$ respectively. $m_j$ and $\rho_j$ denote the mass and density of the $j$-th particle. The kernel function $W(\bm x_i - \bm x_j, h)$ and its derivative $\nabla_{\bm x_i} W(\bm x_i - \bm x_j, h)$ are simplified as $W(\bm x_{ij})$ and $\nabla_i W(\bm x_{ij})$ hereafter.

\subsection{Conventional SPH discretization of governing equations}

In CESPH, the governing equations (\ref{eq1}) and (\ref{eq2}) are discretized in the following forms \cite{bui2008lagrangian}

\begin{equation}\label{con1}
   \dfrac{\mathrm{d} \rho_i}{\mathrm{d}t} = \sum_j m_j \left( \bm v_i - \bm v_j \right)  \nabla_i W(\bm x_{ij})
\end{equation}

\begin{equation}\label{con2}
   \dfrac{\mathrm{d} \bm v_i}{\mathrm{d}t} = \sum_j m_j \left( \dfrac{\bm \sigma_i}{\rho_{i}^2} + \frac{\bm \sigma_j}{\rho_{j}^2} - \pi_{ij} \bm I - p_{ij}^a \bm I \right)  \nabla_i W(\bm x_{ij})
\end{equation}
where $\pi_{ij}$ and $p_{ij}^a$ are the artificial viscosity and the artificial pressure respectively. The artificial viscosity $\pi_{ij}$ is required for stabilizing computations. The following form of $\pi_{ij}$ \cite{monaghan1983shock} is used

\begin{equation}\label{artificial}
    \pi_{ij}= 
\begin{cases}
    \dfrac{-\beta_1 \bar{c}_{ij}\mu_{ij} + \beta_2 \mu^2_{ij}}{\bar{\rho}_{ij}},& \text{if } \bm v_{ij}\cdot\bm x_{ij}\le 0\\
    0,              & \text{otherwise}
\end{cases}
\end{equation} with
\begin{equation}
\mu_{ij} = \dfrac{h \bm v_{ij}\cdot \bm x_{ij}}{\norm{\bm x_{ij}}^2+0.01h^2}
\end{equation}
where $\beta_1$ and $\beta_2$ are two parameters controlling the magnitude of the artificial viscosity, $\bm v_{ij} = \bm v_i - \bm v_j$ is the relative velocity, $c_{ij} = 0.5 (c_i + c_j)$ is the average speed of sound computed as $c_i = \sqrt{E_i/\rho_i}$, with $E_i$ being the elastic modulus of the material. The artificial viscosity should provide stabilized solution without over-damping the system.

For CESPH, the tensile instability is a well-known deficiency which severely limits its application. In slope stability analysis, geomaterials usually have moderate cohesion. It is natural to have some computational areas with tensile stress, where tensile instability develops, which leads to low accuracy or even breakdown of simulations. Therefore, in CESPH, additional numerical stabilization is required. In this work, the widely-used artificial pressure $p_{ij}^a$ is employed to alleviate the tensile instability in CESPH. 

The artificial pressure stabilization adds a positive pressure to prevent particle clumping in the area with tensile stress. It has the following form \cite{monaghan2000sph, gray2001sph}.

\begin{equation}\label{eq_ap}
   p^{a}_{ij}=\gamma \left(\frac{p_i^a}{\rho^2_i}+\frac{p_j^a}{\rho^2_j}\right) \left[\frac{W(\bm x_{ij})}{W(\Delta p)}\right]^n
\end{equation}
where $\gamma$ is a tuning parameter determined through numerical experiments and $n = W(0)/W(\Delta p)$, $\Delta p$ being the initial average particle spacing. $p_i^a$ is taken as $-p_i$ if $p_i < 0$; otherwise, it is set to zero. $p_j^a$ is computed in the same way.

Although artificial pressure can alleviate tensile instability in geotechnical simulations \cite{bui2008lagrangian,peng2015sph,peng2019loquat}, it has two main issues. First, the tuning parameter $\gamma$ needs to be determined in each simulation. Simulations with a different numerical resolution, boundary conditions, and material properties may have different values of the tuning parameter. This is undesired because it is inconvenient and introduces arbitrariness, reducing the credibility of the method. The second issue for the artificial stabilization is that in some instances, it cannot prevent tensile instability even with very large $\gamma$; and further increasing $\gamma$ leads to the breakdown of simulations. Consequently, CESPH cannot be used to solve this kind of problems.

\section{STABILIZED TOTAL-LAGRANGIAN SPH WITH CONFIGURATION UPDATE}

Due to the tensile instability, CESPH may have difficulties in slope stability analysis. It is found that the origin of the tensile instability in CESPH is the employment of Eulerian kernels, which are defined based on the current configuration \cite{belytschko2000unified}. On the other hand, the tensile instability is completely avoided if the kernel functions are computed based on the initial configuration $\bm X$ \citep{belytschko2002stability, rabczuk2004stable}. The initial configuration-based kernels are termed as total-Lagrangian (TL) kernels, and the SPH method employing TL kernels are called total-Lagrangian SPH (TLSPH). Because it does not have tensile instability, it is a suitable method for slope analysis with cohesion. However, the original form of TLSPH cannot model problems with large material shearing and distortion due to the use of initial configuration-based TL kernels. It also has an hourglass instability caused by rank-deficiency \cite{ganzenmuller2015hourglass}. In this section, we first briefly introduce the formulations of TLSPH and propose a configuration update scheme and an hourglass control algorithm to enhance the original TLSPH. These two enhancements enable TLSPH to model problems with large material shearing and distortion without hourglass mode.

\subsection{TLSPH formulations}

Unlike CESPH in which the particle approximation is based on the current particle position $\bm x$, the field variables and its derivative in TLSPH are evaluated using the reference configuration $\bm X$ as

\begin{equation}
   f(\bm X_i) = \sum_j f(\bm X_j) W(\bm X_{ij}) \dfrac{m_j}{\rho_{0j}}
\end{equation}

\begin{equation}
   \nabla f(\bm X_i) = \sum_j f(\bm X_j) \nabla_i W(\bm X_{ij}) \dfrac{m_j}{\rho_{0j}}
\end{equation}
where $W(\bm X_{ij})$ and $\nabla_i W(\bm X_{ij})$ are the total Lagrangian kernel function and its derivative evaluated at the reference configuration $\bm X$. $\rho_{0j}$ is the reference/initial density of the $j$-th particle.

In TLSPH, the conservation Eq. (\ref{eq1}) and (\ref{eq2}) take the following forms \cite{de2013total}:

\begin{equation}\label{ref1}
  \rho = J^{-1} \rho_0
\end{equation}

\begin{equation}\label{ref2}
\dfrac{\mathrm{d} \bm v}{\mathrm{d}t} = \frac{1}{\rho_0} \nabla_0 \cdot \bm{P} + \bm g
\end{equation} 
where the values defined in the reference configuration, $\bm X$ is denoted by a subscript $0$. $J$ is the Jacobian and it is computed based on the deformation gradient matrix $\bm F$ as $J = \mathrm{det}(\bm F)$. $\bm P$ is the first Piola Kirchhoff stress, which is related to the Cauchy stress as 
\begin{equation}
\bm P = J \bm F^{-1} \bm \sigma
\label{eq:StressConversion}
\end{equation} In the above equation, the deformation gradient matrix is computed as

\begin{equation}\label{defmat}
\bm{F} = \dfrac{\mathrm{d}\bm{x}}{\mathrm{d}{\bm{X}}} = \dfrac{\mathrm{d}\bm{u}}{\mathrm{d}{\bm{X}}} + \bm{I}
\end{equation}
where $\bm u = \bm x - \bm X$ represents the displacement vector, and $\bm I$ is an identity matrix.

In TLSPH, only the momentum Eq. (\ref{ref2}) is solved, as the continuity Eq. (\ref{ref1}) is satisfied naturally. The discretized form of Eq. (\ref{ref2}) in a total Lagrangian manner is as follows

\begin{equation}\label{con_tl}
   \dfrac{\mathrm{d} \bm v_i}{\mathrm{d}t} = \sum_j m_j \left( \dfrac{\bm P_i}{\rho_{0i}^2} + \frac{\bm P_j}{\rho_{0j}^2} - \bm{\Pi}_{ij} \right)  \nabla_i W(\bm X_{ij})
\end{equation} 
where $\bm \Pi_{ij} = J \bm F^{-1} \pi_{ij}$ is the pullback of the artificial viscosity. As mentioned before, the TLSPH formulation is free from any tensile instability; hence, the use of artificial pressure $p_{ij}^a$ is not required. 

For geomaterials, the same DP model is used in TLSPH simulations. Therefore, the strain rate tensor and spin tensor are needed. To this end, we first evaluate the velocity gradient tensor $\bm l$
\begin{equation}
\bm l = \dot{\bm F} \bm F^{-1}
\end{equation} where $\dot{\bm F}$ is the rate of deformation gradient obtained as
\begin{equation}
\dot{\bm{F}} = \dfrac{\mathrm{d}\dot{\bm{x}}}{\mathrm{d}{\bm{X}}} = \dfrac{\mathrm{d}\bm{v}}{\mathrm{d}{\bm{X}}}
\end{equation}

The discretized forms of the deformation gradient matrix $\bm F$ and its rate $\dot{\bm F}$ are

\begin{equation}\label{deformation}
   \bm{F}_i = - \sum_j \left(\bm u_i - \bm u_j \right) \otimes \nabla_i W(\bm X_{ij}) \dfrac{m_j}{\rho_{0j}} + \bm I \\
            = - \sum_j \left(\bm x_i - \bm x_j \right) \otimes \nabla_i W(\bm X_{ij}) \dfrac{m_j}{\rho_{0j}}
\end{equation}

\begin{equation}\label{rate_deformation}
   \bm{\dot{F}}_i = - \sum_j \left(\bm v_i - \bm v_j \right) \otimes \nabla_i W(\bm X_{ij}) \dfrac{m_j}{\rho_{0j}}
\end{equation}

With the velocity gradient $\bm l$, the strain rate tensor $\dot{\bm \varepsilon}$ and spin tensor $\bm \omega$ can be computed as
\begin{equation}
\dot{\bm \varepsilon} = \dfrac{1}{2}(\bm l + \bm l^\mathrm{T}),~~~~\bm \omega = \dfrac{1}{2}(\bm l - \bm l^\mathrm{T})
\end{equation}

The integration of the constitutive model in TLSPH is similar to that in CESPH. Once the Cauchy stress is obtained from the constitutive model, it is converted to the first Piola Kirchhoff stress using Eq.~(\ref{eq:StressConversion}), which is then used in the momentum conservation equation (\ref{con_tl}) to compute the acceleration.

\subsection{Update of configuration in TLSPH}\label{up}
The TLSPH in its original form is incapable of modeling plastic flows and post-failure behaviors of geomaterials. This is due to the utilization of the initial/reference configuration $\bm X$ in the computation. Whenever there is severe deformation of the material, e.g. particle penetration or material separation, the Jacobian $J$ becomes negative, leading to unrealistic prediction and breakdown of simulation. The reason behind this is that when the distortion between the reference configuration $\bm X$ and the current configuration $\bm x$ is too large, the deformation gradient $\bm F$ does not remain a well-conditioned one. The ill-conditioned deformation gradient leads to erroneous values of the field variables like stress and strain in the computation.

To solve this problem, Ganzenm{\"u}ller et al. \cite{ganzenmuller2016hourglass} and Leroch et al. \cite{leroch2016smooth} proposed to update the initial/ reference frame periodically to the current configuration 
\begin{equation}
   \bm X \leftarrow \bm x
\end{equation}
\begin{equation}
   V \leftarrow J V
\end{equation}
where the reference particle position and the particle volume $V$ are updated. This update does not change the stress tensor, or deformation status, thus preserves stress and deformation history. After the update, all the computations are based on the new reference configuration using the same TLSPH formulations until the next update. It is found that with the configuration update problems with large material distortion and plastic flow can be modeled using TLSPH \cite{ganzenmuller2016hourglass,leroch2016smooth}.

An essential consideration of the configuration update is how frequently the reference configuration needs to be updated. In \cite{ganzenmuller2016hourglass,leroch2016smooth}, the authors proposed to update the reference configuration when the distance between a pair of interacting particles exceeds one half of the smoothing length. This approach was used in the simulation of a tension test of a hyperelastic patch and the scratching in visco-plastic materials using TLSPH. However, they did not provide much details on how to perform the configuration update and why to choose this specific criterion, which is the key to investigate the performance of the TLSPH method with the configuration update.

As the reference configuration needs to be updated whenever the material distortion is too large, a measurement of distortion is required. In this work, we quantify the distortion in the neighborhood of particle $i$ as the maximal normalized displacement in its support domain
\begin{equation}
   d_i = \max_j \left(\left| \frac{x_{ij} - X_{ij}}{X_{ij}} \right|\right)
\end{equation} This definition can be interpreted as the maximal line strain, with respect to the reference configuration, in the support domain of $i$. At each step, the maximal normalized displacement is computed at each particle, then an overall maximal normalized displacement $d^\mathrm{max}$ is obtained as
\begin{equation}
d^{\max} = \max_i d_i
\end{equation} The overall maximal normalized displacement $d^{\max}$ is employed to quantify the material distortion in the simulation. The reference configuration is updated to the current configuration when the following criterion is met
\begin{equation}
d^{\max} \geq k
\label{eq:UpdateCriterion}
\end{equation} where $k$ is a constant defining the acceptable distortion in the simulation, whose appropriate value and influence on results will be discussed in Section~\ref{sec:NumerucalExamples}. It is found that the criterion with the normalized displacement can more realistically consider the distortion in TLSPH than that used in \cite{ganzenmuller2016hourglass,leroch2016smooth}, where the initial distances between particles are not taken into account.

\subsection{Stabilization with hourglass control}
\label{subsec:HourglassControl}
Being a collocation method having approximation and numerical integration at the same set of particles, SPH suffers from rank deficiency, which usually leads to hourglass mode instability. In CESPH, because particles are often in irregular distribution due to large displacement and distortion, the development of hourglass mode is restrained. Therefore, in CESPH, the problem of hourglass mode is usually negligible, so it does not attract attention. However, in TLSPH, because the computation is based on the reference configuration, where particles generally have regular distribution, the hourglass model develops unboundedly and can eventually lead to poor results or even failure of simulation.

Hourglass mode in TLSPH manifests patterns of particle displacements under which no strain is produced. As a result, no stress can be generated to resist this mode of particle displacement, just like the zero-energy mode in the finite element method with reduced integration. Dyka and Ingel \cite{dyka1995approach} proposed to add additional stress points for strain and stress computations, which can sufficiently suppress hourglass mode. However, the stress point method gives rise to a much higher computational cost. More significantly, there is a lack of detailed investigation on where to place the stress points and how to move them in simulations with large deformation. Consequently, SPH with stress points appears not to find widespread applications.

Because the TLSPH is entirely free from tensile instability, hourglass mode becomes the dominant source of instability in TLSPH simulations. To this end, a stabilization term for hourglass control is imperative for TLSPH. The hourglass control is particularly crucial for long term simulations, as the hourglass mode develops over time \cite{belytschko2000unified}. In this work, an hourglass control algorithm proposed by Ganzenm\"uller \citep{ganzenmuller2015hourglass, ganzenmuller2016hourglass} is employed. This hourglass control algorithm is used in TLSPH simulation of hyperelastic \cite{zhan2019stabilized} and elastoplastic \cite{leroch2016smooth} materials, where accurate results and fast convergence are observed. The basic concept of the hourglass control algorithm is to identify the displacements corresponding to the hourglass mode and then enforce additional forces to minimize these unphysical displacements.

As shown in Eq. (\ref{defmat}), the deformation gradient tensor $\bm F$ for each particle is computed based on the deformation/displacement of its neighboring particles in the support domain. Ideally, the displacement described by this deformation gradient should be a linear function across the support domain. The hourglass mode, therefore, can be identified as the difference between the actual displacements and the displacements corresponding to the deformation gradient. For instance, for a particle, $i$ with reference position $X_i$ and deformation gradient $\bm F_i$, the ideal linear separation between particle $i$ and its neighboring particle $j$ can be written as

\begin{equation}
   \left<\bm x_{ij}\right>^i = \bm F_i \bm X_{ij}
\end{equation} 
where $\bm X_{ij} = \bm X_i - \bm X_j$. However, as mentioned before the actual separation in the current frame is $\bm x_{ij} = \bm x_i - \bm x_j$. The difference between these two linear separations is identified as hourglass mode

\begin{equation}
   \bm e_{ij}^i = \left<\bm x_{ij}\right>^i - \bm x_{ij} 
\end{equation} Due to symmetry for the $j$-th particle, the error of the pair can be obtained as
\begin{equation}
   \bm e_{ij} = \frac{1}{2} \left( \bm e_{ij}^i + \bm e_{ij}^j \right) = \frac{1}{2} \left( \bm F_i + \bm F_j \right)\bm X_{ij} - \bm x_{ij}
\end{equation}

To control this hourglass mode, a penalty force $\bar{\bm f}_{ij}^{HG}$ along $\bm x_{ij}$ between the $i$-th and $j$-th is employed
\begin{equation}
   \bar{\bm f}_{ij}^{HG} = \frac{1}{2} \alpha E_{ij} \frac{\bm e_{ij} \cdot \bm x_{ij}}{\bm X_{ij}^2} \frac{\bm x_{ij}}{\bm x_{ij}^2}
\end{equation}
where $\alpha$ is a penalty parameter taken as 50 \cite{ganzenmuller2015hourglass,ganzenmuller2016hourglass}, $E_{ij}$ is the average of elastic models i.e. $0.5(E_i + E_j)$. For a particle $i$ the total hourglass control force $\bm f_i^{HG}$ can be computed based on averaging over the neighboring particles
\begin{equation}
   \bm f_i^{HG} = V_0^i \sum_j \bar{\bm f}_{ij}^{HG} W(X_{ij}) V_0^j = \frac{1}{2} \frac{m_i}{\rho_{0i}}  \sum_j \alpha E_{ij} \frac{\bm e_{ij} \cdot \bm x_{ij}}{\bm X_{ij}^2} \frac{\bm x_{ij}}{\bm x_{ij}^2} W(X_{ij}) \frac{m_j}{\rho_{0j}}
\end{equation} The hourglass force is added to the momentum Eq. (\ref{con_tl}). The total acceleration with hourglass reads
\begin{equation}
   \dfrac{\mathrm{d} \bm v_i}{\mathrm{d}t} = \dfrac{\mathrm{d} \bm v_i}{\mathrm{d}t}_{\mathrm{Eq}. (\ref{con_tl})} + \frac{\bm f_i^{HG}}{m_i}
\end{equation}

As a summary, Algorithm~\ref{algo1} shows the process of computation in one step using the proposed TLSPH with configuration update and hourglass control. For simplicity, the integration method shown in Table~\ref{algo1} is a simple Euler integration. In actual simulations in this work, a second-order predictor-corrector integration is employed. The details of this integrator can be found in Peng et al. \cite{peng2019loquat}.

\begin{algorithm}[H]\label{algo1}
\caption{The flowchart of the TLSPH with configuration update and hourglass control}
Start: known variables are the reference position $\bm X_i$, the current position $\bm x_i$, velocity $\bm v_i$, and Cauchy stress $\bm \sigma_i$; \\ 
~~~~1. Compute deformation gradient tensor $\bm F_i$ and its rate $\dot{\bm F}_i$ for each particle; \\
~~~~2. Compute the first Piola Kirchhoff stress $\bm P_i$;\\
~~~~3. Obtain total acceleration $\mathrm{d}\bm v_i/\mathrm{d}t$ based on the momentum equation and hourglass control;\\
~~~~4. Update velocity $\bm v_{t+\Delta t}$ and the current position $\bm x_{t+\Delta t}$;\\
~~~~5. Calculate $\dot{\bm \varepsilon}_i$ and $\bm \omega_i$, update Cauchy stress $\bm \sigma_{t+\Delta t}$ using the DP model;\\
~~~~6. Check Eq. (\ref{eq:UpdateCriterion}). If necessary, update the reference configuration and the neighbor list;\\
~~~~7. Add time $t + \Delta t$.\\
End
\end{algorithm}

\section{Numerical examples}
\label{sec:NumerucalExamples}
The presented stabilized TLSPH method with configuration update is implemented in the GPU-accelerated open-source SPH code LOQUAT \cite{peng2019loquat} and is employed to model problems with large deformation and failure in this section. First, the collapse of a column of cohesive soil and its post failure movement are modelled using CESPH and TLSPH. Results from these two methods are compared. The influence of the configuration update in TLSPH and the optimal criterion for updating are investigated. The influence of the hourglass control algorithm is also discussed. Next, the TLSPH is used to evaluate the safety factor of a slope. The results are compared with those from CESPH.

\begin{figure}[hbtp!]
\centering
\includegraphics[width=0.4\textwidth]{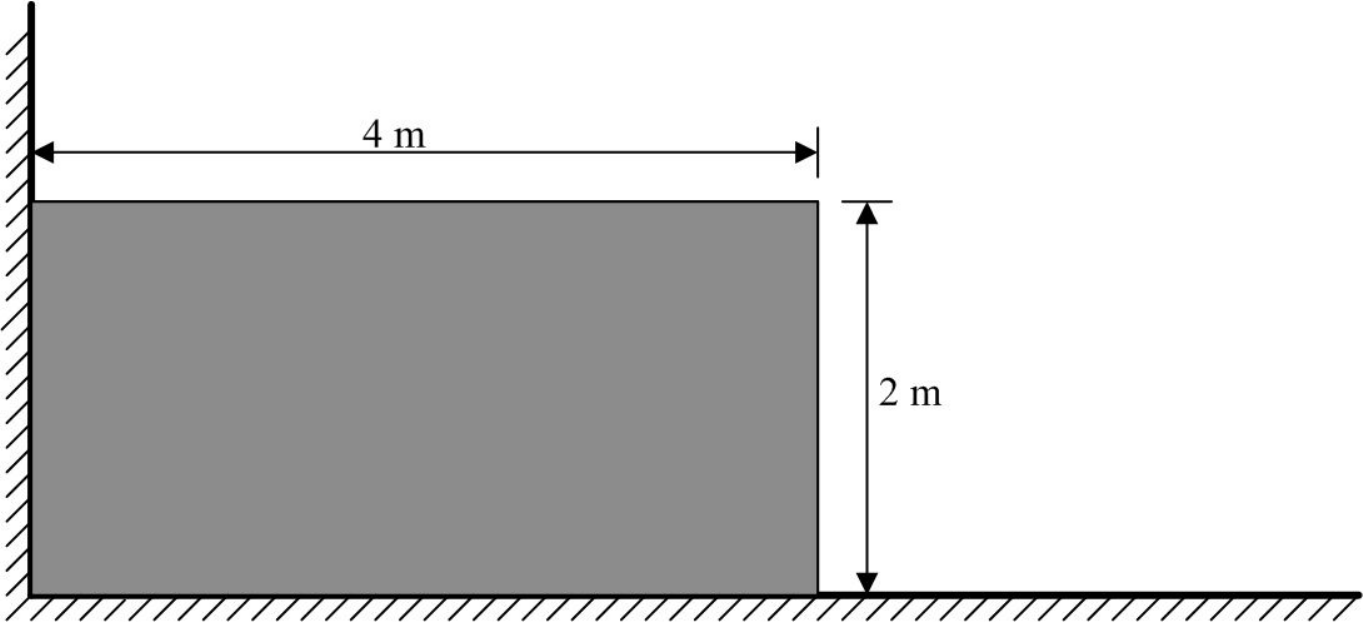} 
\caption{Setup of the soil block for large deformation analysis}\label{soil_block}
\end{figure} 

\begin{table}[h!]
\caption{Material properties of cohesive soil}\label{soil_coltab}
\centering
\begin{tabular}{llll}
\hline
Density ($\rho$)  & Young's module ($E$) & Friction angle ($\phi$)  & Cohesion ($c$)     \\ \hline
1850~kg/m$^3$     & 1.5~MPa            & 25$^\circ$               & 5 KPa  \\ \hline                                                             
\end{tabular}
\end{table}
 
\begin{table}[h!]
\caption{Computational parameters for the soil collapse simulation}\label{soil_col}
\centering
\begin{tabular}{llll}
\hline
Particle spacing ($\Delta p$)    & Smoothing length ($h$)   & Time step ($\Delta t$) & Artificial viscosity parameters ($\beta_1,~\beta_2$) \\ \hline
0.03 m                           & 0.045 m                  & $1 \times 10^{-4}$ s   & (2.5, 2.5)                                                \\ \hline
\end{tabular}
\end{table}

\begin{figure}[hbtp!]
\centering
\includegraphics[width=0.65\textwidth]{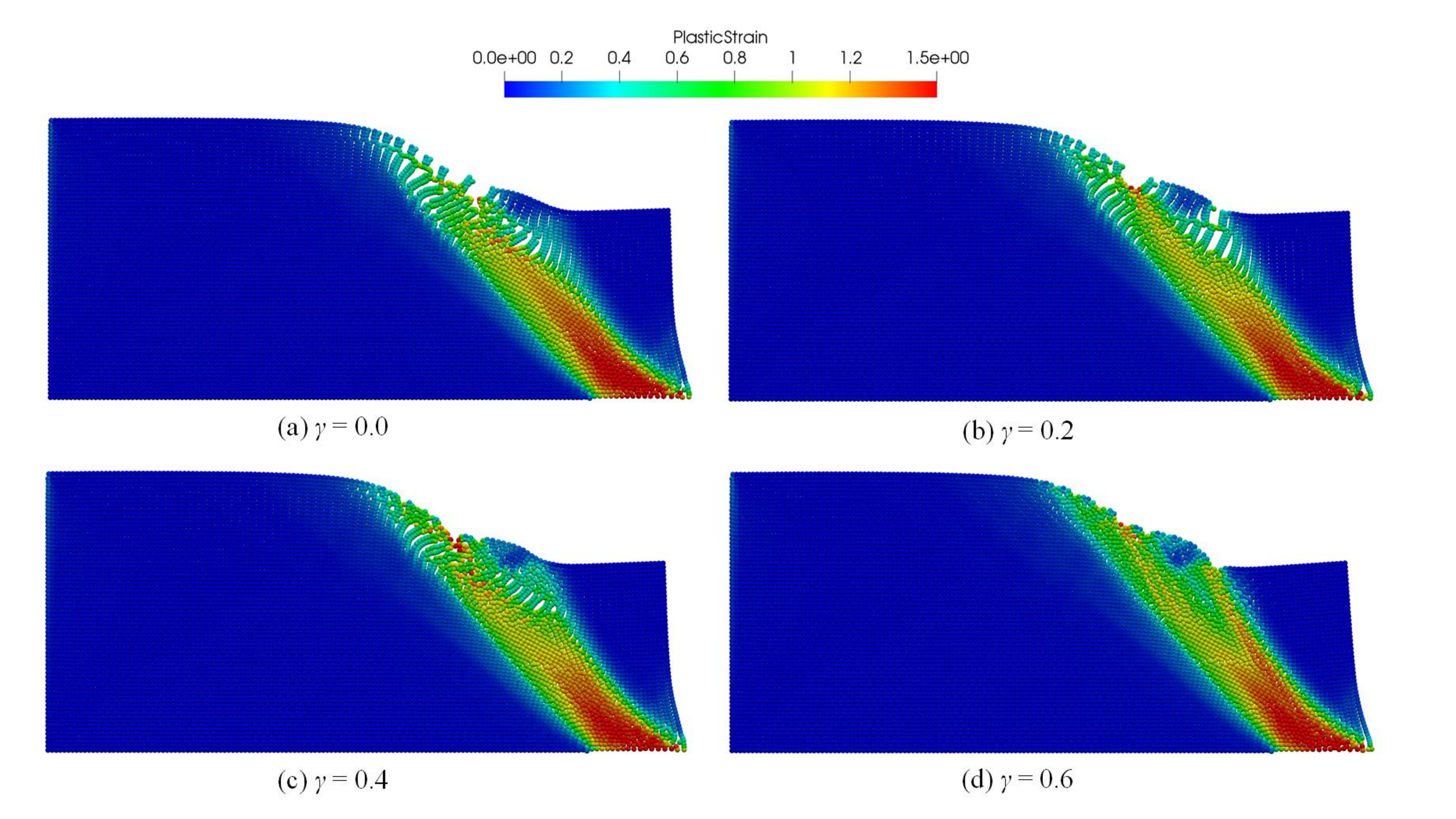} 
\caption{Conventional SPH and removal of tensile instability with different artificial pressure $\gamma$ parameter}\label{instability}
\end{figure} 

\subsection{Large deformation and failure of cohesive soil}
This test highlights the accuracy and stability of the present TLSPH framework in modeling large deformation and post-failure flows of cohesive soils. As shown in Fig. \ref{soil_block}, a rectangular column of cohesive soil subjected to gravity is considered. The density of the soil particles is $1850$ kg/m$^3$. The material parameters and other computational variables are provided in Table \ref{soil_coltab} and \ref{soil_col}. The DP model is used to simulate the soil. The total simulated time is 6 seconds.

\subsubsection{Tensile instability}
As mentioned before, conventional SPH suffers from tensile instability, which should be alleviated using the artificial pressure method. However, the constant $\gamma$ in the artificial pressure method needs to be calibrated for each simulation with different numerical settings and material properties. For instance, when simulating elastic material, Gray et al. \cite{gray2001sph} found that $\gamma = 0.3$ is the optimal choice. On the other hand, Bui et al. \cite{bui2008lagrangian} found that $\gamma = 0.5$ produced best results in SPH simulation of geomaterials.

\begin{figure}[hbtp!]
\centering
\includegraphics[width=0.65\textwidth]{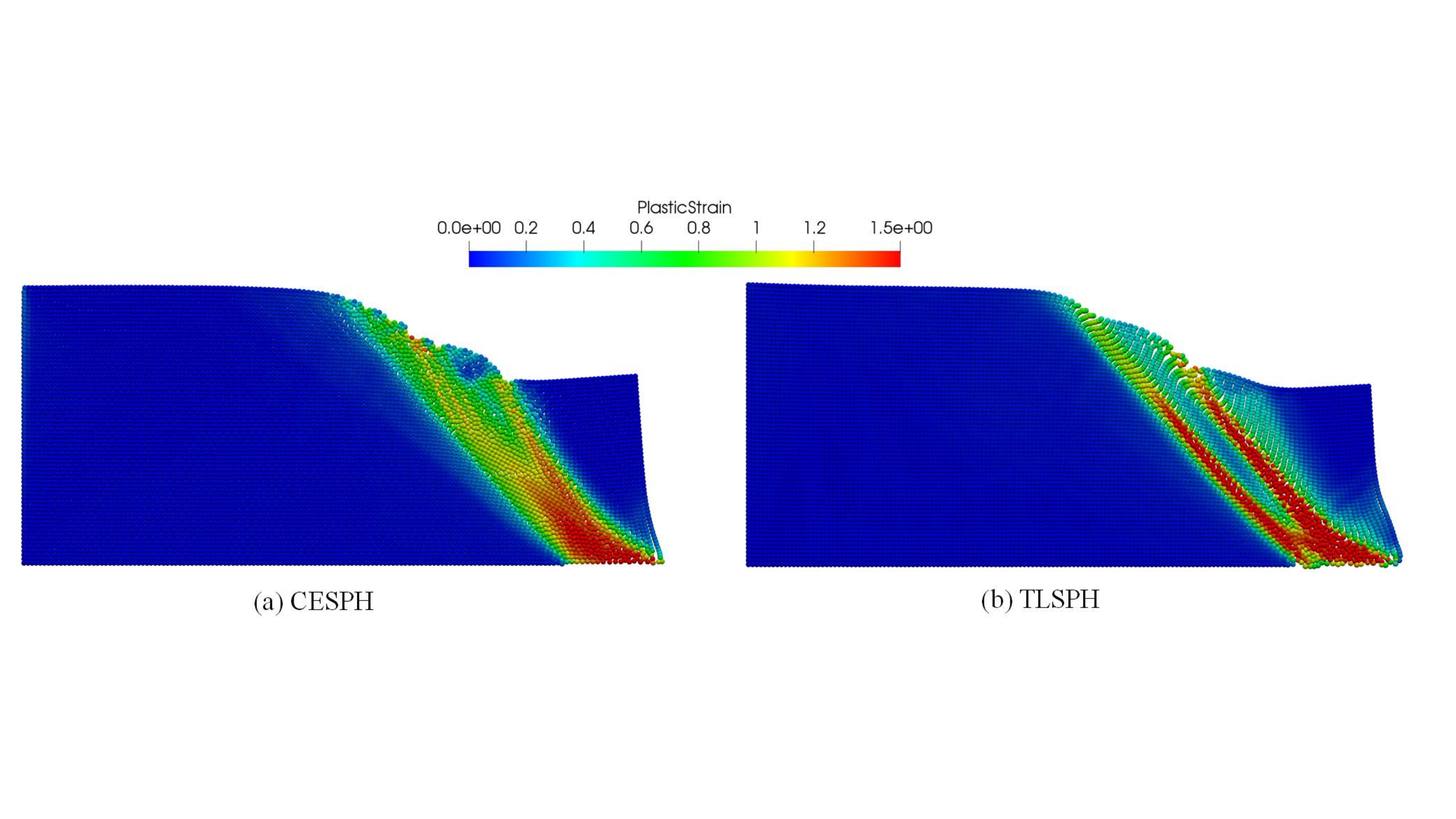}
\caption{Final deformed shape free from tensile instability with conventional SPH ($\gamma = 0.6$) and TLSPH}\label{instability_comp}
\end{figure} 
 
In this work, four CESPH simulations with varying $\gamma$ are carried out. The results of the deformed soil body and the distribution of accumulated plastic strain are shown in Figure~\ref{instability}. It is obvious that when no treatment is used ($\gamma = 0.0$), the results suffer from tensile instability in the form of unphysical cracking and particle clumping. The results are improved with $\gamma = 0.2$ and 0.4, but cracking can still be observed. The tensile instability is completely removed in the simulation with $\gamma = 0.6$. Further increasing $\gamma$ leads to the breakdown of the simulation. It is found that with tensile instability, CESPH could not correctly model the failure pattern, which features a branching in the shear band. This branching of the shear band can only be observed if tensile instability is sufficiently suppressed, as observed in the results from $\gamma=0.6$.

\begin{figure}[hbtp!]
\centering
\includegraphics[width=0.9\textwidth,trim={20 330 30 0}, clip]{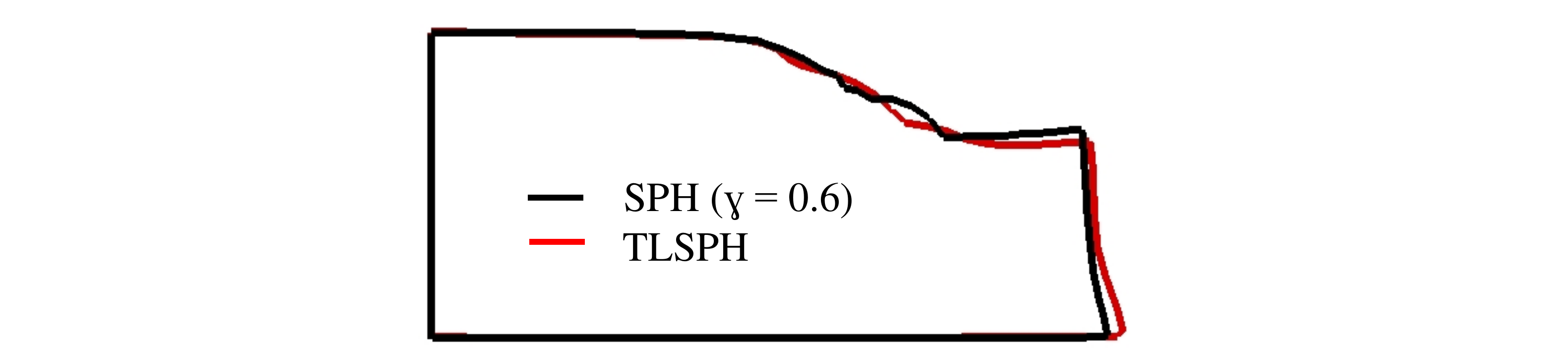} 
\caption{Comparison of the stabilised deformed shape with conventional SPH ($\gamma = 0.6$) and TLSPH}\label{soil_shape}
\end{figure} 

\begin{figure}[hbtp!]
\centering
\includegraphics[width=0.65\textwidth]{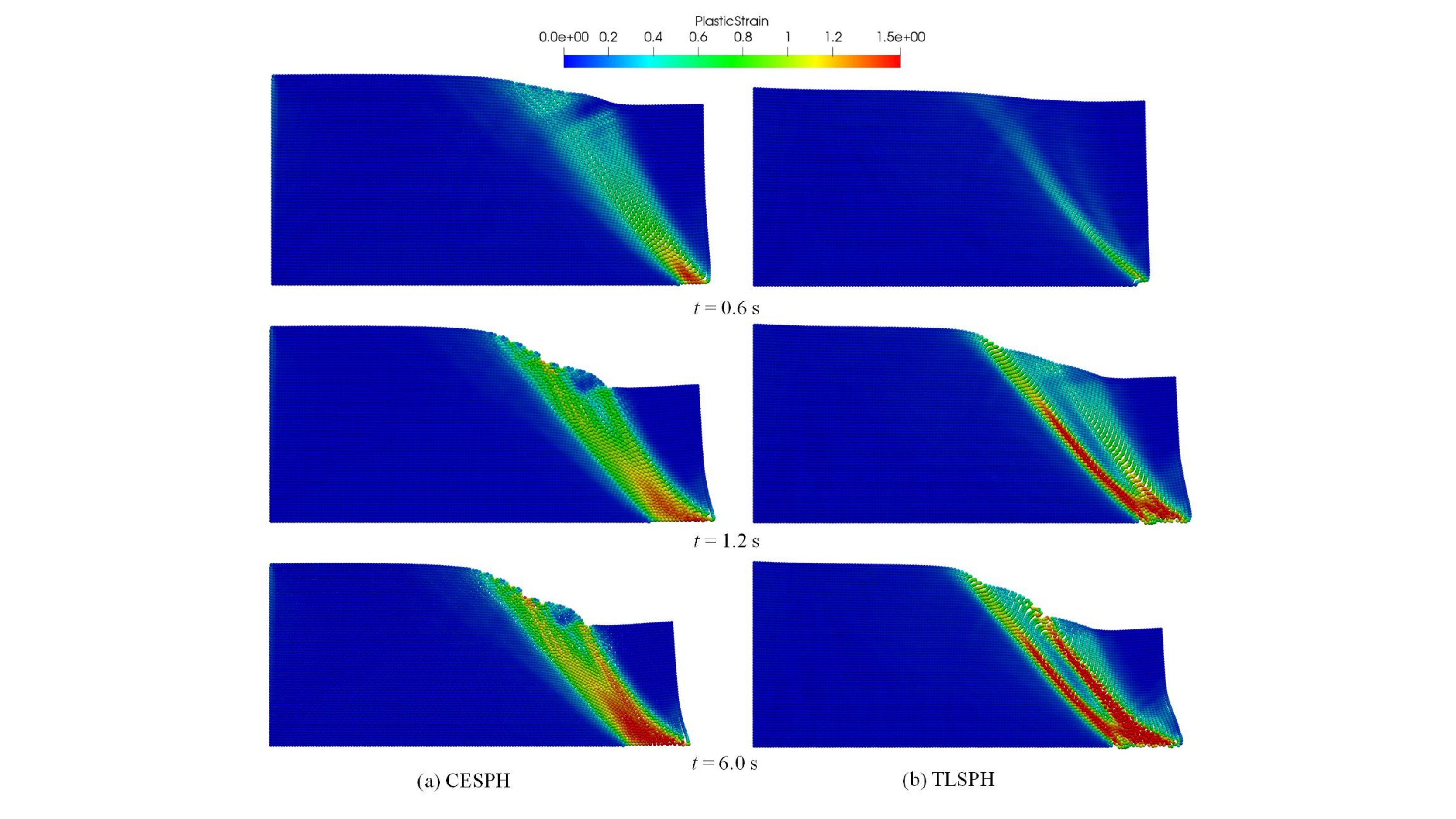} %trim={<left> <lower> <right> <upper>}
\caption{Comparison of the deformed shape and accumulated plastic strain at different time step with conventional SPH ($\gamma = 0.6$) and TLSPH}\label{soil_pl_comp}
\end{figure}

The same case is simulated using the proposed TLSPH with configuration update. The comparison between the results from CESPH and TLSPH is shown in Figure~\ref{soil_pl_comp}, where the artificial pressure constant for CESPH is $\gamma=0.6$, and the configuration update criterion for TLSPH is $k=2$. It is observed that with the configuration update, the proposed TLSPH can model problems with large deformation and post-failure flow. It does not need tensile instability treatment and overcomes the difficulty of the original TLSPH in large deformation modeling. Moreover, the presented TLSPH produces results with two distinct shear bands, giving more clear modeling of the progressive failure of the soil column. On the other hand, the shear band in the CESPH results is smeared out on a broader area, indicating that CESPH cannot give detailed modeling of the failure process. This is particularly problematic if one wants to investigate the failure mechanism in more complex cases. The final profiles obtained from the two methods are shown in Figure~\ref{soil_shape}. It is found that the final deformed shapes of the soil body from the two simulations are similar but not without differences. The TLSPH results show a clear step-wise top surface due to the two shear bands. On the other hand, the surface obtained from the CESPH simulation has unphysical bumps caused by artificial pressure.

The deforming and failing process of the soil body obtained using the CESPH and TLSPH are shown in Figure~\ref{soil_pl_comp}. A shear band first initiates at the toe of the column and then propagates into the soil. In the results from CESPH, the shear band is smeared into a wider area, and a branching results in a secondary shear band near the top surface. On the other hand, the second shear band is formed at an early stage in the simulation using TLSPH. Two distinct shear bands are obtained in the TLSPH results. It is found that the TLSPH with configuration update is free of tensile instability and can model large deformation and failure of geomaterials sufficiently. The CESPH, although giving rise to stable simulations with tuned tensile instability coefficients, cannot accurately capture the complete failure process.

\subsubsection{Update of configurations in TLSPH}
\begin{figure}[hbtp!]
\centering
\includegraphics[width=0.65\textwidth]{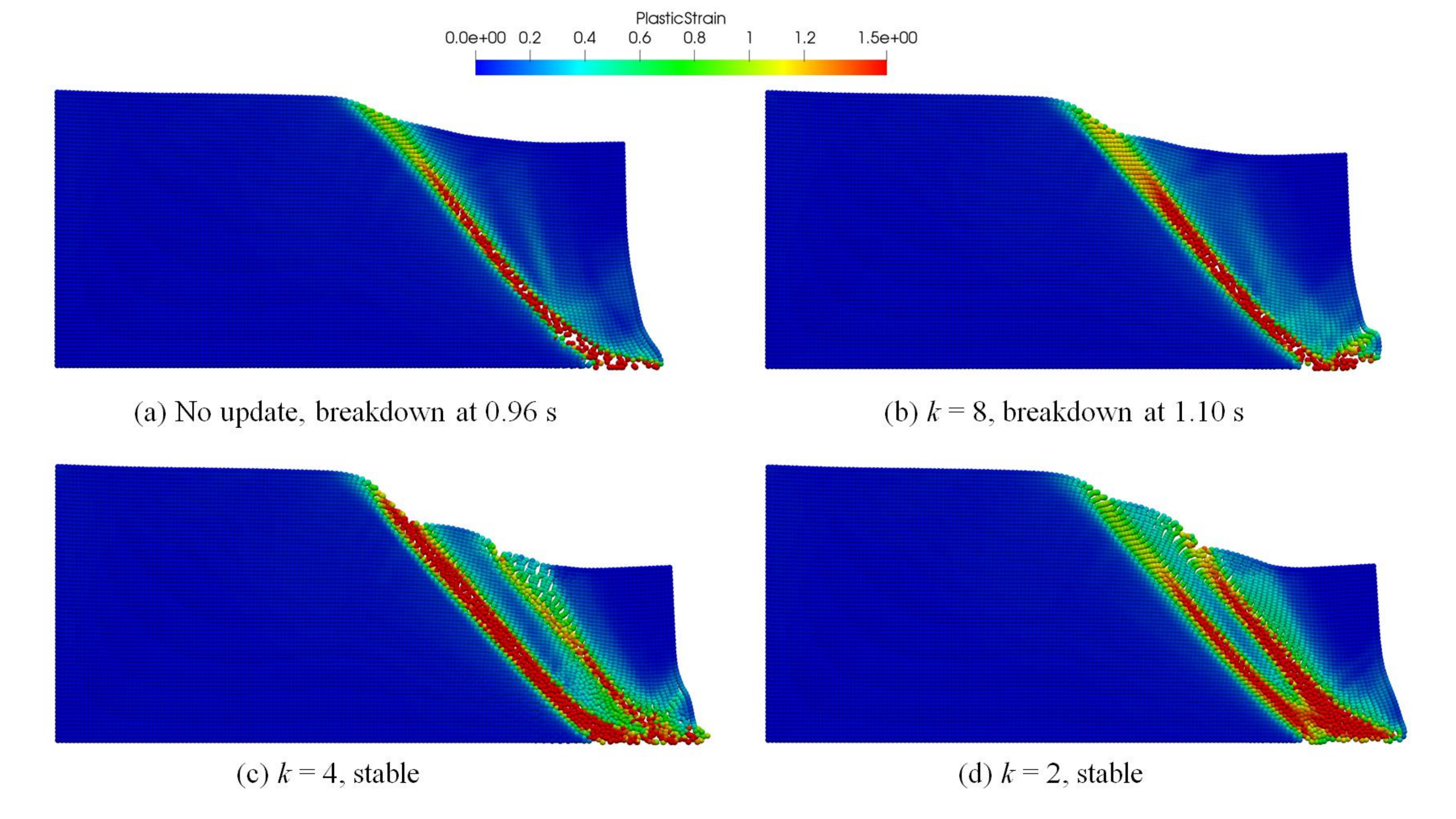} 
\caption{Effect of update of configurations in TLSPH}\label{soil_ref}
\end{figure}

As mentioned in Section \ref{up}, the TLSPH in its original form cannot model large plastic deformation, which is common in geomaterials. For instance, Fig.~\ref{soil_ref} shows the results of the same cohesive soil collapse using TLSPH without configuration update. The simulation terminates at $0.96$ second. This is because the current configuration is much distorted compared with the reference configuration, resulting in unphysical deformation gradient and negative Jacobian at some particles. Therefore, it is necessary to update the reference configuration after a certain degree of distortion. In this work, the overall maximal normalized displacement $d^{\max}$ is employed to measure the distortion. If $d^{\max}$ is larger than a threshold $k$, the reference configuration is updated to the current configuration.  That is, the update is performed if the current inter-particle distance is $k$ times larger than the original inter-particle distance for at least one pair of particles in the computation.

To investigate the influence of the threshold $k$, three simulations using TLSPH are performed with $k = 2$, 4, and 8. The results are also given in Fig.~\ref{soil_ref}. With $k=8$, the simulation terminates at 1.1 s, indicating that the frequency of the configuration update is too low to prevent ill-conditioned deformation gradient. The simulation with $k=4$ is stable, as observed in Fig.~\ref{soil_ref}. However, instabilities can be found in the plastic zones, which means that the deformation gradients are still ill-conditioned. Although this ill-condition does not directly lead to termination of the simulation, it reduces the computational accuracy and results in unphysical deformation patterns. With $k=2$, the simulation is stable and gives rise to well-defined shear bands. No instability is observed. Therefore, $k= 2$ is employed in the following simulations in this work. 

It should be pointed out that the choice of the threshold $k$ can be linked to the physical properties of geomaterials. During shearing of geomaterials, a physical particle is subjected to the frictional contacts and cohesion forces from surrounding particles; thus, the movements of the surrounding particles contribute to the macroscopic variables such as strain and stress. However, as the shear deformation develops, these links cease to exist during shearing, and new particles get in contact. Therefore, in numerical simulation, new reference configuration needs to be established. This is the physical foundation of the configuration update. However, the precise frequency of configuration update depends on many factors such as grain size, deformation characteristics, boundary conditions, and numerical resolution. Hence, it is difficult to determine the threshold $k$ quantitatively using physical properties and numerical conditions. Therefore, in this work, $k$ is considered as a numerical constant.

\subsubsection{Hourglass control}
\begin{figure}[hbtp!]
\centering
\includegraphics[width=0.65\textwidth]{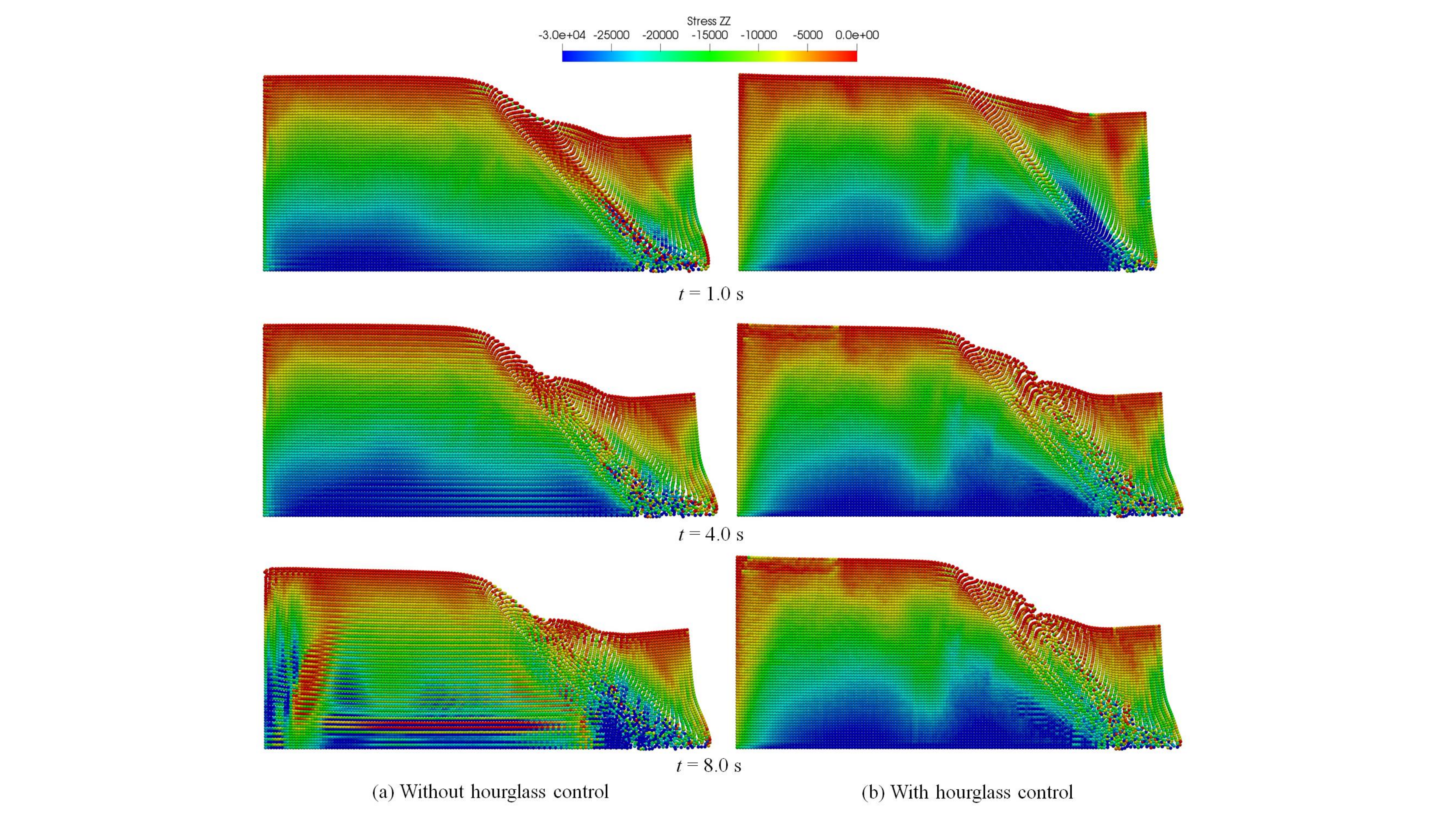}
\caption{Comparison of the effect of hourglass control force at time step with TLSPH}\label{soil_hourglass}
\end{figure}

\begin{figure}[hbtp!]
\centering
\includegraphics[width=0.8\textwidth,trim={100 325 100 0}, clip]{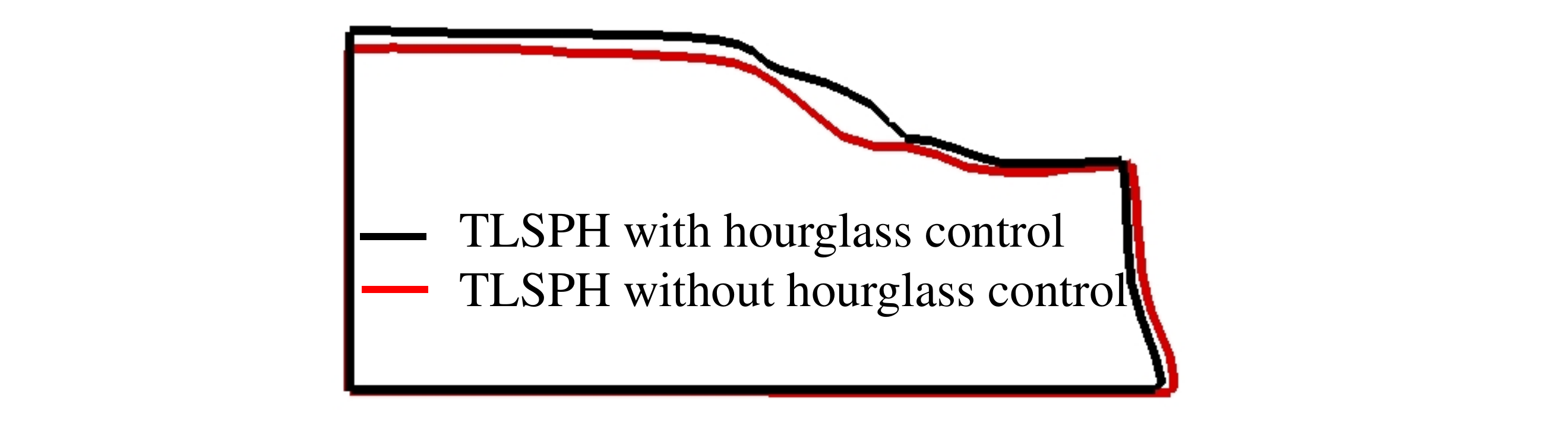} 
\caption{Comparison of the stabilised deformed shape in TLSPH with and without hourglass control}\label{soil_shape_def}
\end{figure}

Previous SPH simulation of large deformation in geomaterials usually did not focus on stress accuracy. In the presented TLSPH, smooth stress results can be obtained if the hourglass control is employed, as demonstrated in Fig.~\ref{soil_hourglass}. It is observed that without hourglass control, the results show typical hourglass instability, having spurious oscillation of stress over adjacent particle layers. The hourglass instability develops over time, eventually leading to completely distorted particle distribution and poor stress results. More significantly, the distorted particle distribution results in incorrect deformation pattern, as shown in Fig.~\ref{soil_shape_def}. Due to the hourglass instability, the soil body has unphysical settlements, resulting in an erroneous final profile of the deformed soil block. These unphysical settlements are avoided if the hourglass control is activated. It is found that with hourglass control, the simulation is stable even in long term computations. Moreover, the stress results are smooth; no spurious oscillations can be observed even after a long time of computation. As accurate modeling of stress is crucial in certain applications, e.g. soil-structure interactions, it is a desirable feature for the presented TLSPH method to provide satisfactory stress results. 

\subsection{Safety factor analysis}
In this section, the TLSPH framework is utilised to evaluate the factor of safety of a homogeneous slope. The strength reduction approach \cite{griffiths1999slope} is applied to calculate the safety factor. In this approach, the cohesion $c$ and friction angle $\phi$ of the soil is reduced incrementally by a factor as
\begin{equation}
   c_r = \frac{c}{f_s}
\end{equation}

\begin{equation}
   \phi_r = \frac{\phi}{f_s}
\end{equation} The reduced values of $c_r$ and $\phi_r$ are then used to check the stability of the slope. The largest value at which the failure of the slope occurs is considered the factor of safety. Once the factor of safety is determined, the post-failure large deformation of the slope is obtained by continuing the simulation. 

\begin{figure}[hbtp!]
\centering
\includegraphics[width=0.6\textwidth]{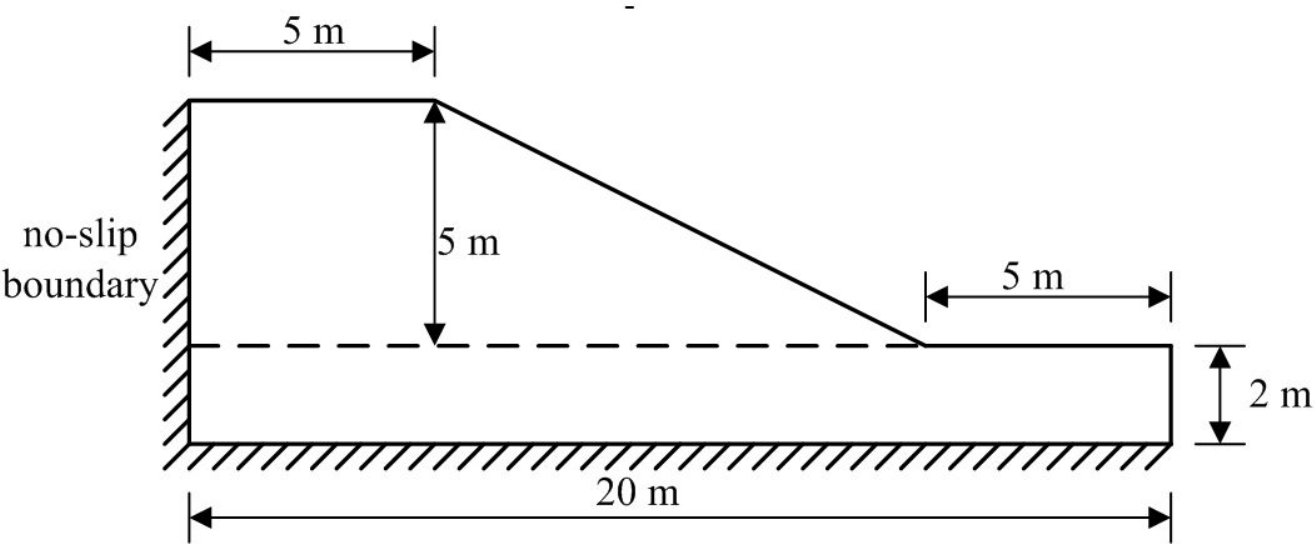} 
\caption{Setup of the slope for factor of safety analysis}\label{soil_slope}
\end{figure} 

\begin{table}[h!]
\caption{Material properties of soil for the slope failure set up}\label{slope_coltab}
\centering
\begin{tabular}{llll}
\hline
Density ($\rho$)  & Young's module ($E$) & Friction angle ($\phi$)  & Cohesion ($c$)     \\ \hline
1850~kg/m$^3$     & 100.0~MPa            & 30$^\circ$               & 5 KPa  \\ \hline                                                             
\end{tabular}
\end{table}
 
\begin{table}[h!]
\caption{Computational parameters for the slope failure simulation}\label{slope_col}
\centering
\begin{tabular}{llll}
\hline
Particle spacing ($\Delta p$)    & Smoothing length ($h$)   & Time step ($\Delta t$) & Artificial viscosity parameters ($\beta_1,~\beta_2$) \\ \hline
0.1 m                           & 0.15 m                  & $1 \times 10^{-5}$ s   & (2.5, 2.5)                                                \\ \hline
\end{tabular}
\end{table}

The set up of the slope is shown in Fig. \ref{soil_slope}, which has the same geometry and boundary condition as in \cite{peng2019loquat}. The criterion of slope failure is based on the development of the maximum displacement in the slope. This criterion was first proposed by Bui et al. \cite{bui2011slope} for CESPH analysis of slope stability. Initially, the slope is in an equilibrium state with the original material constants. After the sudden strength reduction, deformations develop in the slope. If the slope remains stable under a reduction factor, the displacement is only caused by the disturbance in material constants. As a result, the maximum displacement should be small and become stable in a short time, and the shape of the time-maximum displacement curve is convex. On the other hand, if the slope cannot remain stable, the maximum displacement increases quickly, and the curve is concave. In this work, the factor $f_s$ is initially taken as $1.0$ and then increased by $0.1$ for each simulation. The maximum displacement for each simulation is recorded over time. If under a certain factor, the time-maximum displacement curve is concave, this factor is taken as the safety factor of the slope. 

\begin{figure}[hbtp!]
\centering
\includegraphics[width=0.75\textwidth]{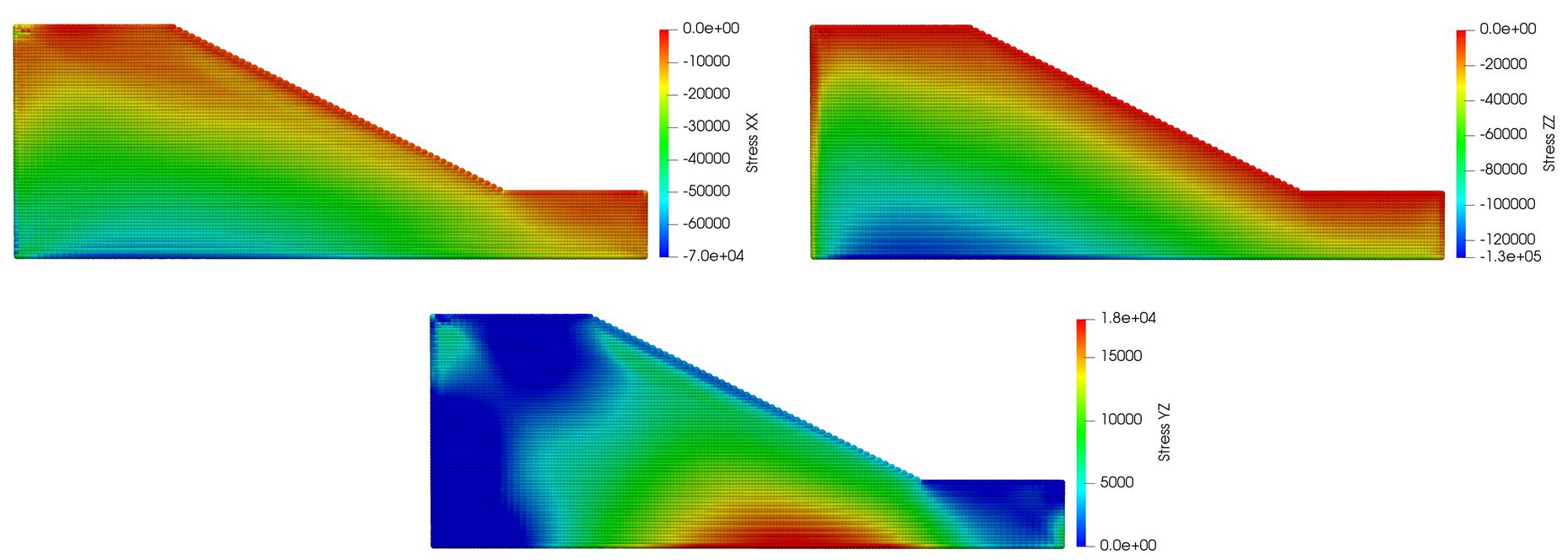} 
\caption{Distribution of initial stress}\label{soil_stress}
\end{figure} 

\begin{figure}[hbtp!]
\centering
\begin{subfigure}[t]{0.45\textwidth}    %trim={<left> <lower> <right> <upper>}
\includegraphics[width=\textwidth]{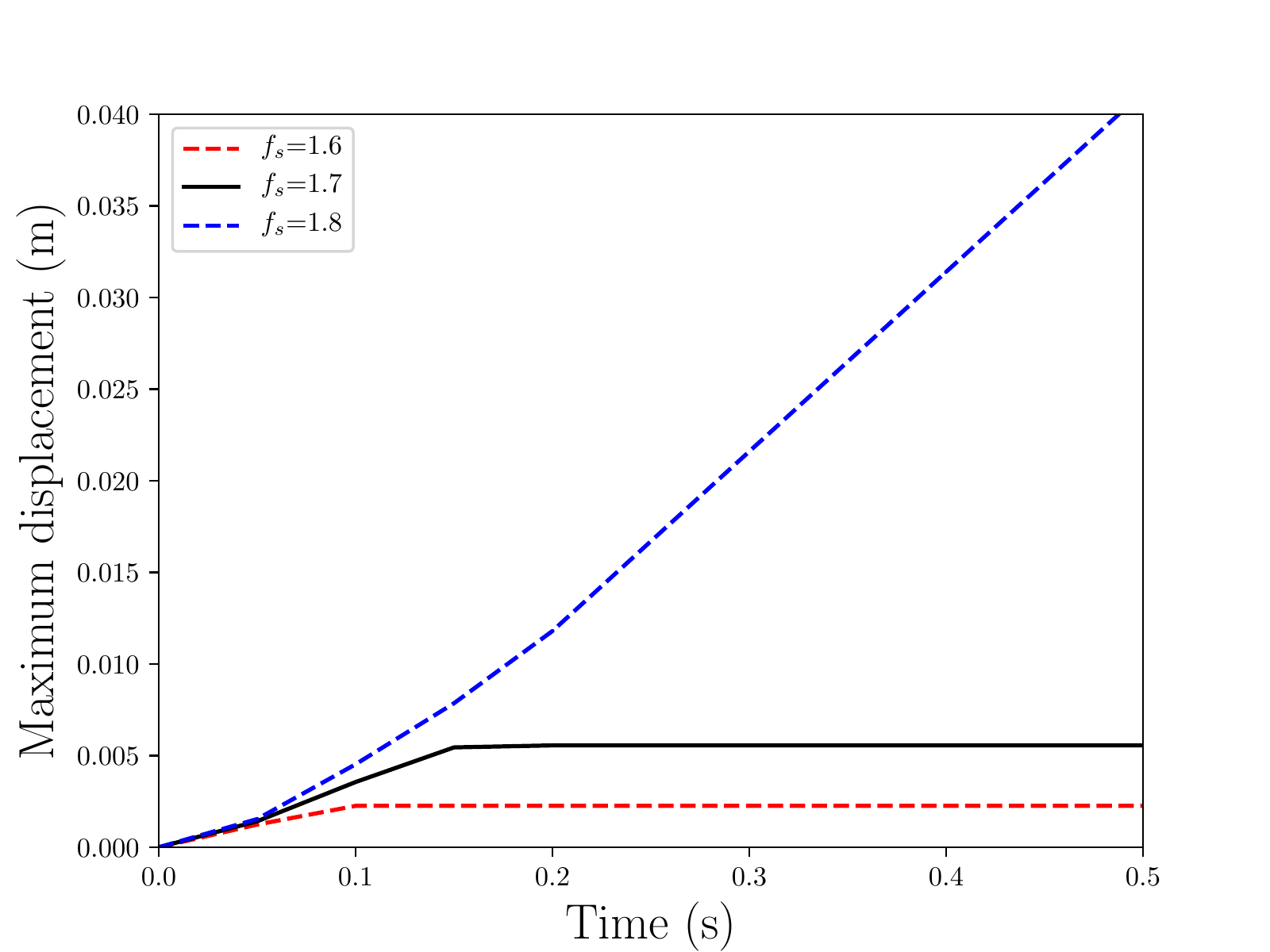}
\caption{Conventional SPH}\label{kal1}
\end{subfigure}
\begin{subfigure}[t]{0.45\textwidth}
\includegraphics[width=\textwidth]{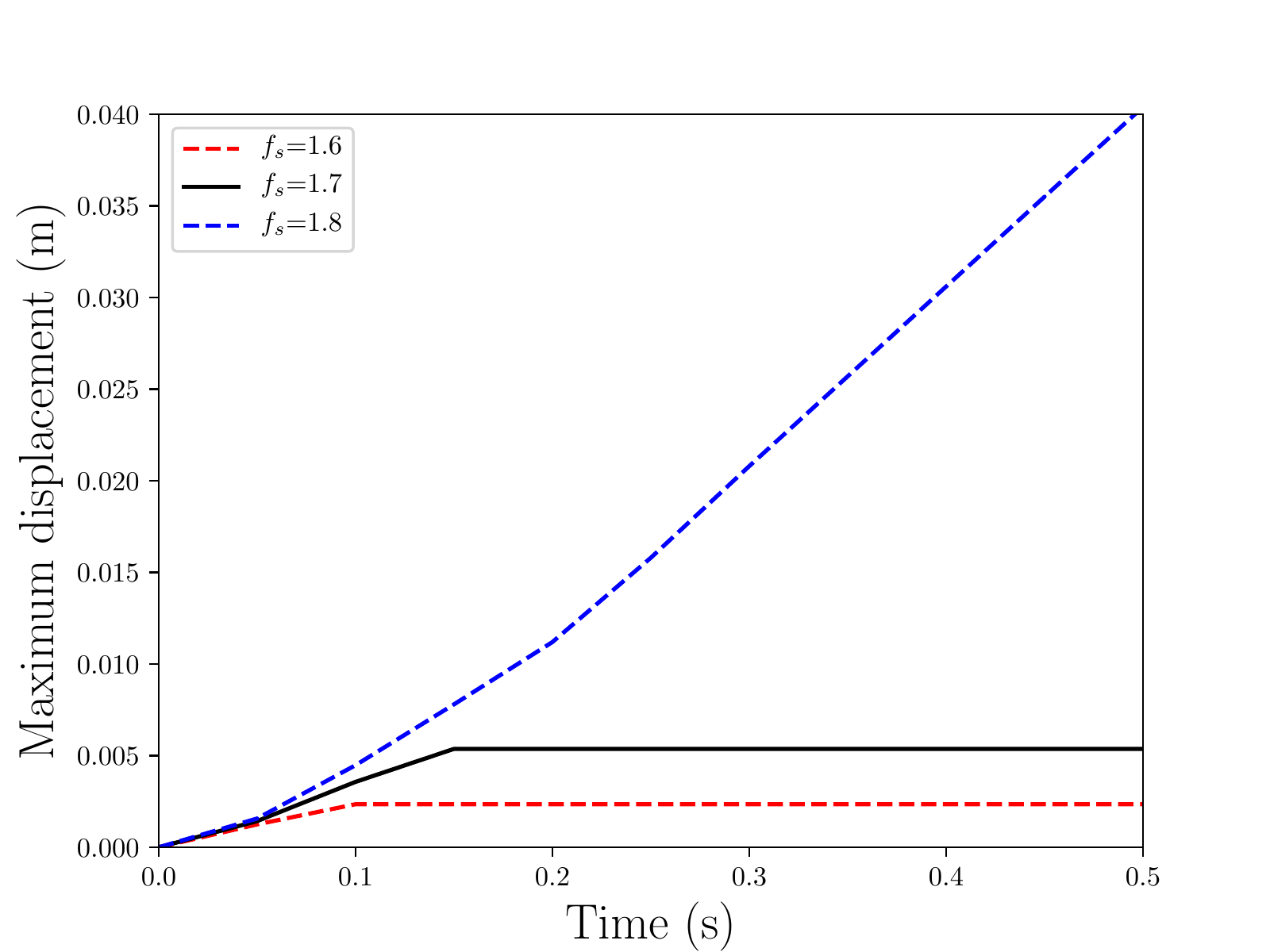}
\caption{TLSPH}\label{kal2}
\end{subfigure}
\caption{Displacement over time at different factor of safety}\label{fas}
\end{figure}

The material constants and the numerical settings are given in Table~\ref{slope_coltab} and \ref{slope_col}, respectively. Initially, the original material constants are employed in the computation to get the initial geostatic stresses. The numerical damping approach introduced by Bui et al. \cite{bui2013improved} is used to accelerate the consolidation process. Once the steady state is achieved, the original particle positions are restored, and all variable except stresses are set as zero. Then the reduced material constants are employed in the simulation to check the stability of the slope. The obtained initial stresses in the slope are given in Fig.~\ref{soil_stress}. It is found using the TLSPH with hourglass control, the accuracy of the stress fields is satisfactory. 

\begin{figure}[hbtp!]
\centering
\includegraphics[width=0.65\textwidth]{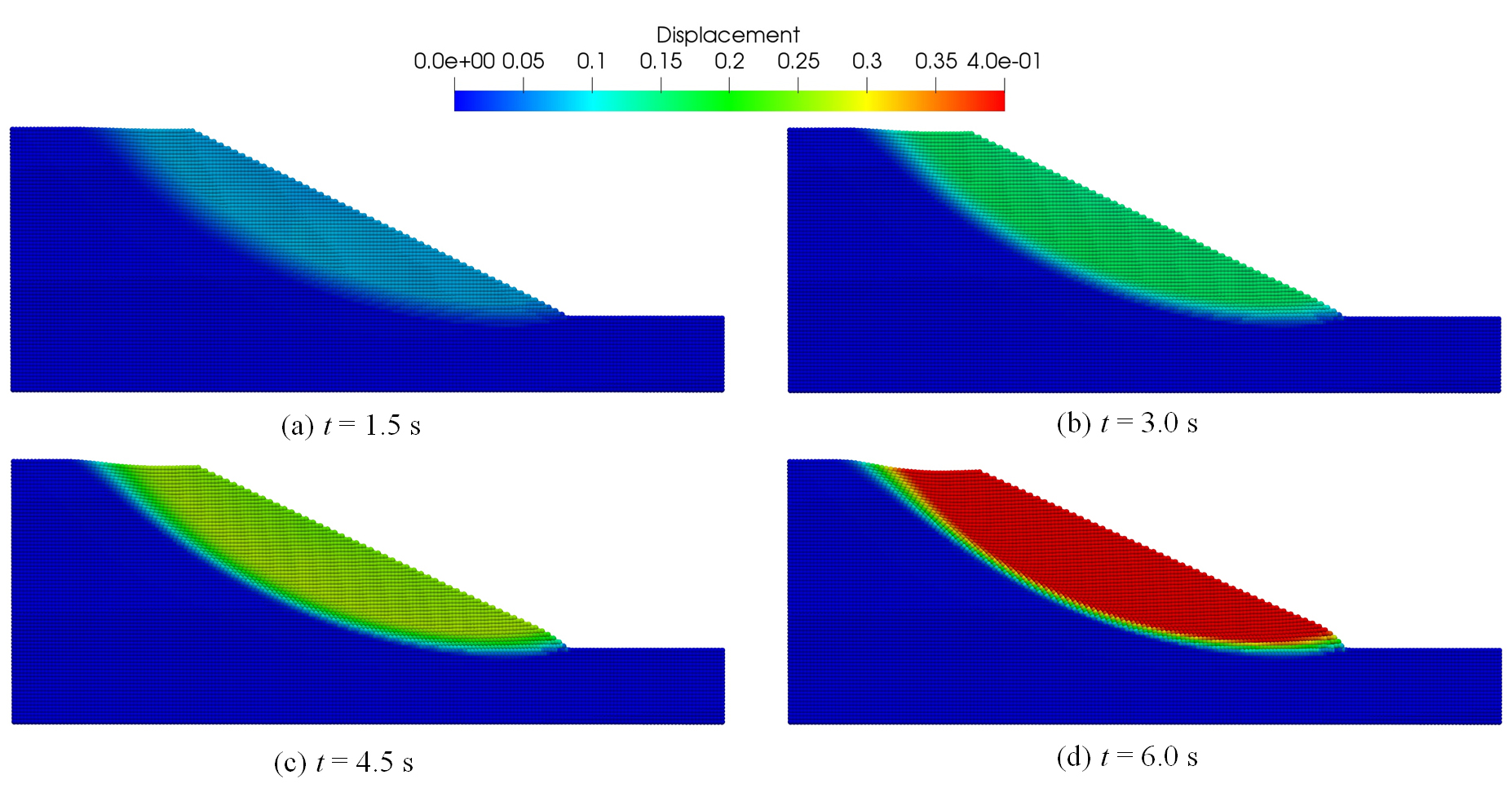} 
\caption{Displacement of the slope at different time steps with $f_s=1.8$}\label{slope_disp}
\end{figure} 

\begin{figure}[hbtp!]
\centering
\includegraphics[width=0.65\textwidth]{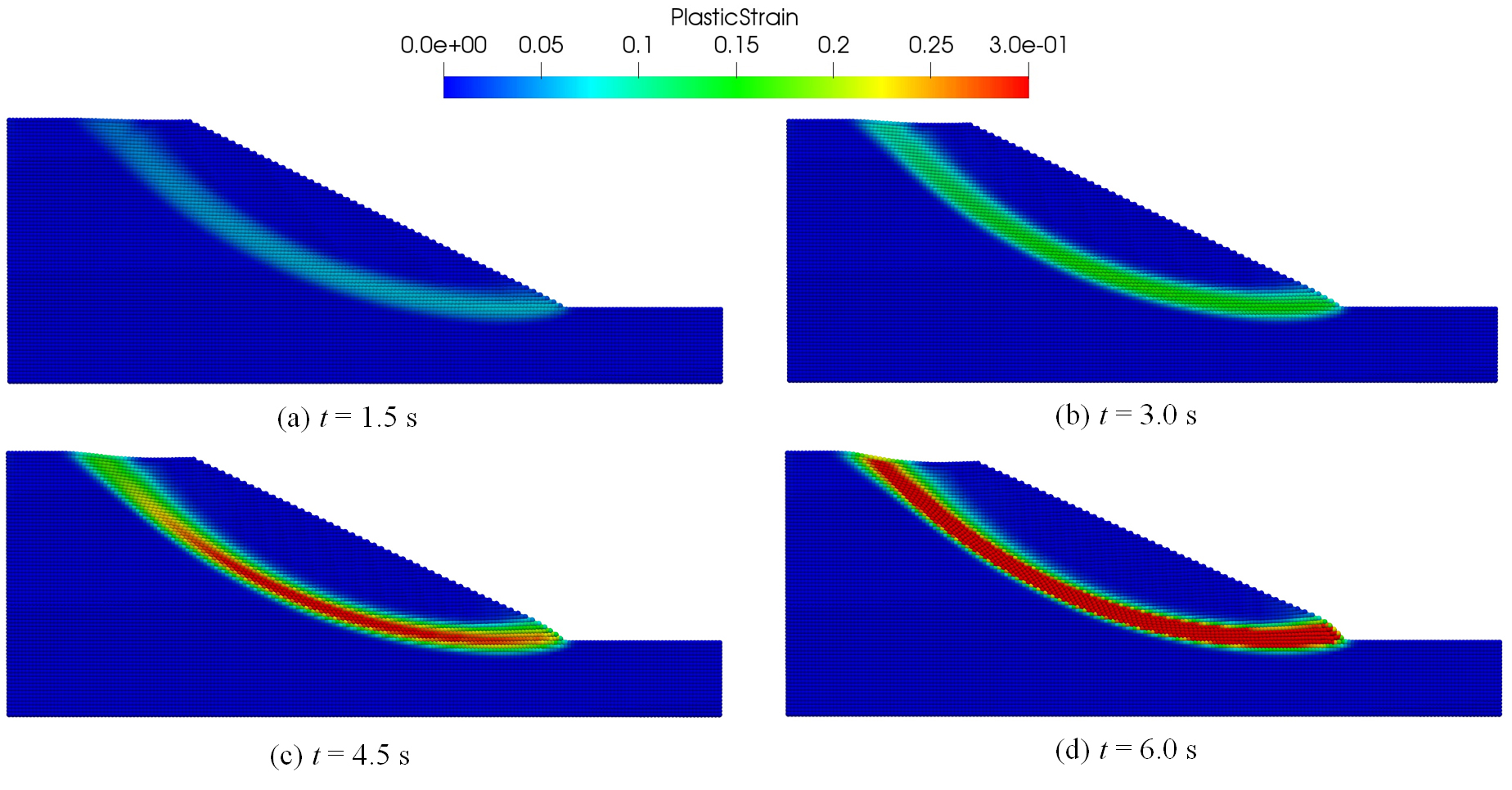} 
\caption{Plastic strain of the slope at different time steps with $f_s=1.8$}\label{slope_pl}
\end{figure} 

Figure~\ref{fas}(b) gives the time history of the maximum displacement under varying reduction factor. It is found that if the reduction factor is less than 1.7, the maximum displacement quickly becomes stable and remains a very small value in the whole computation. Correspondingly, the shape of the time-maximum displacement curve is convex, indicating that the slope is stable if the reduction factor is less than 1.7. On the other hand, the maximum displacement develops unboundedly in the interested time span and reaches very large value. The shape of the curve is concave. Consequently, it is found that the slope becomes unstable with reduction factor 1.8. Therefore, for the TLSPH simulation, the obtained safety factor for this slope is 1.8. Additionally, computations using CESPH are performed, and the results are shown in Fig. \ref{fas}(b). It is observed that the time-maximum displacement curves obtained from the two methods are almost identical. Furthermore, the results of the safety factor are well corroborated by numerical results from literature \cite{peng2019loquat}. 

\begin{figure}[hbtp!]
\centering
\includegraphics[width=0.65\textwidth]{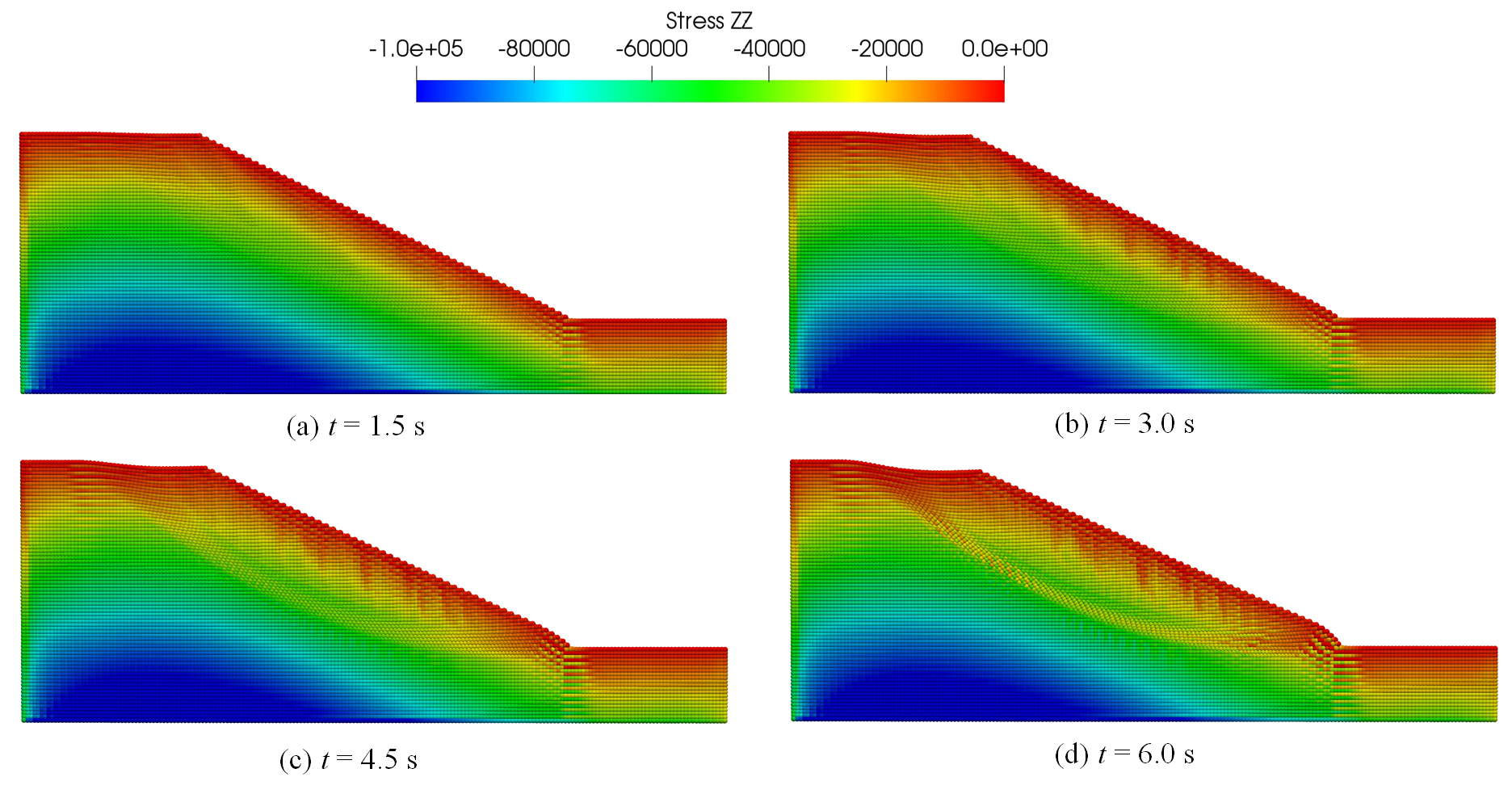} 
\caption{Stress ($\sigma_{zz}$) distribution of the slope at different time steps with $f_s=1.8$}\label{slope_stress}
\end{figure}

The developments of displacement and shear band of the slope under reduction factor $f=1.8$ are given in Fig. \ref{slope_disp} and \ref{slope_pl}, respectively. It is observed that the failure propagates from the toe to the crest of the slope. A well-developed shear band is obtained, whose location and shape are consistent with other numerical results obtained with conventional SPH \cite{bui2011slope,bui2013improved,peng2019loquat}. Above the shear band, a part of the slope body falls and rotates until a new equilibrium state is reached, resulting in concentrated displacements in the sliding body. Although artificial pressure is not used in the TLSPH simulation, no tensile instability can be observed. Furthermore, the distribution of vertical stress at four-time instances are given in Fig.~\ref{slope_stress}. It is found that, except the sliding body, in most part of the slope, the stress does not change. The stress field remains smooth despite the long simulation time. Hourglass instability is avoided, and satisfactory stress results can be obtained, which is a distinct advantage of the presented TLSPH method.

\begin{figure}[hbtp!] 
\centering
\includegraphics[width=0.7\textwidth]{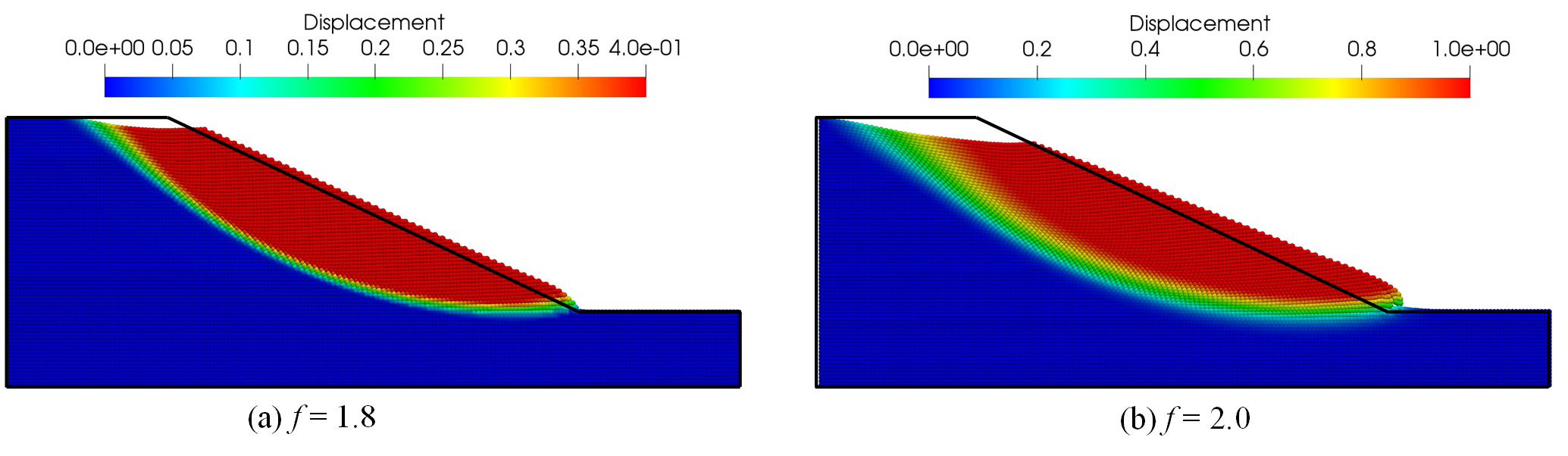} 
\caption{Comparison of the initial configuration (black line) and the deformed slope after $6$ s: (a) $f=1.8$, (b) $f=2.0$.}\label{slope_shape}
\end{figure}

With the configuration update, TLSPH can sufficiently model the post-failure large deformation. Fig.~\ref{slope_shape} gives the final shape of the slope under reduction factor $f=1.8$ and 2.0. Reasonably, a higher reduction factor results in larger post-failure deformation. In both cases, large shear deformations are observed along the sliding surface, where particle distortion is high. The original TLSPH is incapable of modeling this problem. Again, in the simulation with $f=2.0$, no tensile instability or hourglass mode can be observed, demonstrating the high accuracy and robustness of the presented TLSPH method.

\section{CONCLUSION}
In this work, TLSPH is employed to model large deformation and failure in geomaterials. To improve its accuracy and stability, and to make it be able to simulate problems with large deformation, the following enhancements are used: (1) a stiffness-based hourglass control algorithm is applied to circumvent the hourglass mode caused by rank-deficiency; (2) the reference configuration is periodically updated to the current configuration to prevent numerical instability. A criterion on when to perform configuration update is proposed. An elastoplastic constitutive model is incorporated into the stabilized TLSPH to model geomaterials.

Two cases with cohesive soils involving large deformation and plastic flow are modeled using the stabilized TLSPH method.  The results from CESPH and TLSPH show that both methods result in similar deformation patterns provided that appropriate artificial pressure/stress is used in CESPH. However, it is found that TLSPH is free of tensile instability, and it can give more distinct shear bands and more detailed deformation pattern. The criterion on configuration update is discussed, and the optimal choice of the criterion is found to be $k=2$. The hourglass control is found to be essential for the numerical stability in long-term simulations. With the hourglass control, the stabilized TLSPH can give very accurate and smoothed stress fields, which is otherwise difficult for other SPH variants for solid mechanics. The presented stabilized TLSPH proves to be a sufficient numerical tool for large deformation analysis in geomaterials. 

\section*{Acknowledgement}
This work receives funding from the European Union’s Horizon 2020 programme under grant agreement No. 778627 and the Austrian Research Promotion Agency (FFG) under the project No. 865963.

\bibliographystyle{elsarticle-num} 

\begin{thebibliography}{10}
\expandafter\ifx\csname url\endcsname\relax
  \def\url#1{\texttt{#1}}\fi
\expandafter\ifx\csname urlprefix\endcsname\relax\def\urlprefix{URL }\fi
\expandafter\ifx\csname href\endcsname\relax
  \def\href#1#2{#2} \def\path#1{#1}\fi

\bibitem{monaghan2012smoothed}
J.~J. Monaghan, Smoothed particle hydrodynamics and its diverse applications,
  Annual Review of Fluid Mechanics 44 (2012) 323--346.

\bibitem{chen2017meshfree}
J.~S. Chen, M.~Hillman, S.~W. Chi, Meshfree methods: progress made after 20
  years, Journal of Engineering Mechanics 143~(4) (2017) 04017001.

\bibitem{bardenhagen2004generalized}
S.~G. Bardenhagen, E.~M. Kober, The generalized interpolation material point
  method, Computer Modeling in Engineering and Sciences 5~(6) (2004) 477--496.

\bibitem{maeda2006development}
K.~Maeda, H.~Sakai, M.~Sakai, Development of seepage failure analysis method of
  ground with smoothed particle hydrodynamics, JSCE Structural
  Engineering/Earthquake Engineering 23~(2) (2006) 307s--319s.

\bibitem{naili20052d}
M.~Naili, T.~Matsushima, Y.~Yamada, A {2D} smoothed particle hydrodynamics
  method for liquefaction induced lateral spreading analysis, JSCE Journal of
  Applied Mechanics 8 (2005) 591--599.

\bibitem{bui2008lagrangian}
H.~H. Bui, R.~Fukagawa, K.~Sako, S.~Ohno, Lagrangian meshfree particles method
  ({SPH}) for large deformation and failure flows of geomaterial using
  elastic--plastic soil constitutive model, International Journal for Numerical
  and Analytical Methods in Geomechanics 32~(12) (2008) 1537--1570.

\bibitem{bui2011slope}
H.~H. Bui, R.~Fukagawa, K.~Sako, J.~C. Wells, Slope stability analysis and
  discontinuous slope failure simulation by elasto-plastic smoothed particle
  hydrodynamics {(SPH)}, G{\'e}otechnique 61~(7) (2011) 565--574.

\bibitem{bui2013improved}
H.~H. Bui, R.~Fukagawa, An improved {SPH} method for saturated soils and its
  application to investigate the mechanisms of embankment failure: Case of
  hydrostatic pore-water pressure, International Journal for Numerical and
  Analytical Methods in Geomechanics 37~(1) (2013) 31--50.

\bibitem{bui2014novel}
H.~H. Bui, J.~K. Kodikara, A.~Bouazza, A.~Haque, P.~G. Ranjith, A novel
  computational approach for large deformation and post-failure analyses of
  segmental retaining wall systems, International Journal for Numerical and
  Analytical Methods in Geomechanics 38~(13) (2014) 1321--1340.

\bibitem{chambon2011numerical}
G.~Chambon, R.~Bouvarel, D.~Laigle, M.~Naaim, Numerical simulations of granular
  free-surface flows using smoothed particle hydrodynamics, Journal of
  Non-Newtonian Fluid Mechanics 166~(12-13) (2011) 698--712.

\bibitem{peng2016unified}
C.~Peng, X.~G. Guo, W.~Wu, Y.~Q. Wang, Unified modelling of granular media with
  smoothed particle hydrodynamics, Acta Geotechnica 11~(6) (2016) 1231--1247.

\bibitem{wang2014frictional}
J.~Wang, D.~Chan, Frictional contact algorithms in {SPH} for the simulation of
  soil--structure interaction, International Journal for Numerical and
  Analytical Methods in Geomechanics 38~(7) (2014) 747--770.

\bibitem{zhan2019three}
L.~Zhan, C.~Peng, B.~Y. Zhang, W.~Wu, Three-dimensional modeling of granular
  flow impact on rigid and deformable structures, Computers and Geotechnics 112
  (2019) 257--271.

\bibitem{ganzenmuller2016hourglass}
G.~C. Ganzenm{\"u}ller, M.~Sauer, M.~May, S.~Hiermaier, Hourglass control for
  smooth particle hydrodynamics removes tensile and rank-deficiency
  instabilities, The European Physical Journal Special Topics 225~(2) (2016)
  385--395.

\bibitem{swegle1995smoothed}
J.~Swegle, D.~Hicks, S.~Attaway, Smoothed particle hydrodynamics stability
  analysis, Journal of Computational Physics 116~(1) (1995) 123--134.

\bibitem{dyka1997stress}
C.~T. Dyka, P.~W. Randles, R.~P. Ingel, Stress points for tension instability
  in {SPH}, International Journal for Numerical Methods in Engineering 40~(13)
  (1997) 2325--2341.

\bibitem{monaghan2000sph}
J.~J. Monaghan, {SPH} without a tensile instability, Journal of Computational
  Physics 159~(2) (2000) 290--311.

\bibitem{gray2001sph}
J.~P. Gray, J.~J. Monaghan, R.~P. Swift, {SPH} elastic dynamics, Computer
  Methods in Applied Mechanics and Engineering 190~(49-50) (2001) 6641--6662.

\bibitem{peng2019loquat}
C.~Peng, S.~Wang, W.~Wu, H.~S. Yu, C.~Wang, J.~Y. Chen, {LOQUAT}: an
  open-source {GPU}-accelerated {SPH} solver for geotechnical modeling, Acta
  Geotechnica (2019) 1--19.

\bibitem{belytschko2000unified}
T.~Belytschko, Y.~Guo, W.~Kam~Liu, S.~Ping~Xiao, A unified stability analysis
  of meshless particle methods, International Journal for Numerical Methods in
  Engineering 48~(9) (2000) 1359--1400.

\bibitem{bonet2002alternative}
J.~Bonet, S.~Kulasegaram, Alternative total {L}agrangian formulations for
  corrected smooth particle hydrodynamics ({CSPH}) methods in large strain
  dynamic problems, Revue Europ{\'e}enne des {\'E}l{\'e}ments Finis 11~(7-8)
  (2002) 893--912.

\bibitem{vignjevic2006sph}
R.~Vignjevic, J.~R. Reveles, J.~Campbell, {SPH} in a total {L}agrangian
  formalism, Computer Modeling in Engineering \& Sciences 14~(3) (2006) 181.

\bibitem{ganzenmuller2015hourglass}
G.~C. Ganzenm{\"u}ller, An hourglass control algorithm for {Lagrangian} smooth
  particle hydrodynamics, Computer Methods in Applied Mechanics and Engineering
  286 (2015) 87--106.

\bibitem{zhan2019stabilized}
L.~Zhan, C.~Peng, B.~Zhang, W.~Wu, A stabilized {TL}--{WC} {SPH} approach with
  {GPU} acceleration for three--dimensional fluid--structure interaction,
  Journal of Fluids and Structures 86 (2019) 329--353.

\bibitem{wendland1995piecewise}
H.~Wendland, Piecewise polynomial, positive definite and compactly supported
  radial functions of minimal degree, Advances in computational Mathematics
  4~(1) (1995) 389--396.

\bibitem{monaghan1983shock}
J.~Monaghan, R.~A. Gingold, Shock simulation by the particle method sph,
  Journal of Computational Physics 52~(2) (1983) 374--389.

\bibitem{peng2015sph}
C.~Peng, W.~Wu, H.~S. Yu, C.~Wang, A {SPH} approach for large deformation
  analysis with hypoplastic constitutive model, Acta Geotechnica 10~(6) (2015)
  703--717.

\bibitem{belytschko2002stability}
T.~Belytschko, S.~Xiao, Stability analysis of particle methods with corrected
  derivatives, Computers \& Mathematics with Applications 43~(3-5) (2002)
  329--350.

\bibitem{rabczuk2004stable}
T.~Rabczuk, T.~Belytschko, S.~Xiao, Stable particle methods based on
  {L}agrangian kernels, Computer Methods in Applied Mechanics and Engineering
  193~(12-14) (2004) 1035--1063.

\bibitem{de2013total}
T.~De~Vuyst, R.~Vignjevic, Total lagrangian sph modelling of necking and
  fracture in electromagnetically driven rings, International Journal of
  Fracture 180~(1) (2013) 53--70.

\bibitem{leroch2016smooth}
S.~Leroch, M.~Varga, S.~Eder, A.~Vernes, M.~R. Ripoll, G.~Ganzenm{\"u}ller,
  Smooth particle hydrodynamics simulation of damage induced by a spherical
  indenter scratching a viscoplastic material, International Journal of Solids
  and Structures 81 (2016) 188--202.

\bibitem{dyka1995approach}
C.~Dyka, R.~Ingel, An approach for tension instability in smoothed particle
  hydrodynamics (sph), Computers \& structures 57~(4) (1995) 573--580.

\bibitem{griffiths1999slope}
D.~V. Griffiths, P.~A. Lane, Slope stability analysis by finite elements,
  Geotechnique 49~(3) (1999) 387--403.

\end{thebibliography}
\biboptions{sort&compress}


%\begin{thebibliography}{00}
%
%%% \bibitem[Author(year)]{label}
%%% Text of bibliographic item
%\end{thebibliography}
\end{document}